\documentclass[times,final]{elsarticle}

\usepackage{jcomp}
\usepackage{framed,multirow}

\usepackage{amssymb}
\usepackage{latexsym}
\usepackage[fleqn]{amsmath}

\usepackage{url}
\usepackage{xcolor}
\usepackage{ulem}
\newcommand{\etal}{{\it et al.}}
\definecolor{newcolor}{rgb}{.8,.349,.1}

\journal{Journal of Computational Physics}

\begin{document}
\newcommand{\sgn}{\textrm{sgn}}
\newcommand{\bu}{\mathbf{u}}

\verso{Wang \etal}

\begin{frontmatter}

\title{Discrete exterior calculus discretization of two-phase incompressible Navier-Stokes equations with a conservative phase field method}%

\author[1]{Minmiao Wang}
\ead{minmiao.wang@kaust.edu.sa}
\author[1]{Pankaj Jagad}
\ead{pankaj.jagad@kaust.edu.sa}
\author[2]{Anil N. Hirani}
\ead{hirani@illinois.edu}
\author[1]{Ravi Samtaney\corref{cor1}}
\ead{ravi.samtaney@kaust.edu.sa}
\cortext[cor1]{Corresponding author}

\address[1]{Mechanical Engineering, Physical Science and Engineering Division, King Abdullah University of Science and Technology, Jeddah, Saudi Arabia}
\address[2]{University of Illinois at Urbana-Champaign, Department of Mathematics, 1409 W Green St, Urbana, IL 61801, U. S. A}

\received{XXX}
\finalform{XXX}
\accepted{XXX}
\availableonline{XXX}
\communicated{XXX}

\begin{abstract}
We present a discrete exterior calculus (DEC) based discretization scheme for incompressible two-phase flows. Our physically-compatible exterior calculus discretization of single phase flow is extended to simulate immiscible two-phase flows with discontinuous changes in fluid properties such as density and viscosity across the interface. The two-phase incompressible Navier-Stokes equations and conservative phase field equation for interface capturing are first transformed into the exterior calculus framework. The discrete counter part of these smooth equations is obtained by substituting with discrete differential forms and discrete exterior operators. We prove the boundedness of the method for the first order Euler forward and predictor-corrector time integration schemes  in the DEC framework. With a proper choice of two free parameters, the scheme remains phase field bounded without the requirement of any ad hoc mass redistribution.  We verify the scheme against several standard test cases (for interface capturing) comprising not only the flat domains but also the curved domains, leveraging the advantage that DEC operators are independent of the coordinate system. The results show excellent properties of boundedness, mass conservation and convergence. Moreover, we demonstrate the ability of the scheme towards handling large density and viscosity ratios  as well as surface tension in the simulation of various two phase flow physical phenomena on flat or curved surfaces.
\end{abstract}



\end{frontmatter}


\section{Introduction}
Multiphase flows containing at least two phases exist in numerous industrial applications and are also ubiquitous in our natural environment. In a mixture of two or more immiscible phases, the equilibrium state of the mixture is
separated by a thin region referred to as ``interface". Fluid properties such as density and viscosity are discontinuous  (macroscopically) across the interface.  Moreover, on the interface, there exists a balance between the pressure jump, normal viscous stress jump and surface tension. The discontinuous properties across the interface and singular surface tension make discretizing the multiphase Navier-Stokes equations challenging. Also, one of the core tasks in multiphase flow simulation is providing an accurate interface location.

In multiphase flow simulations, two classes of methods of treating the interface exist: $(1)$ interface capturing, wherein the interface front is represented by a set of connected marker points, and $(2)$ interface tracking, in which a marker function is employed to identify each fluid phase. In the interface tracking method \citep{unverdi1992front}, the connected marker points are advected with the flow and  a specialized implementation is required to restructure the front in order to reflect any topological changes of the interface. On the other hand, interface capturing methods employ an additional equation governing the evolution of marker function, and any change in the topology of the interface occurs automatically, i.e. without any specialized reconstruction. Due to this advantage, interface capturing methods, such as volume of fluid (VOF) \citep{hirt1981volume, youngs1982time, harvie2000new, scardovelli2003interface, pilliod2004second, jofre20143} and the level set (LS) \citep{sussman1994level, olsson2005conservative, olsson2007conservative, herrmann2008balanced, waclawczyk2015consistent} methods are more popular and widely applied in multiphase flow simulations. We note that changes in topology in interface tracking methods are still somewhat limited by or dictated by the mesh resolution used in the simulations. One striking feature of VOF method is that it can conserve mass, while its main drawback is the complexity of interface reconstruction. In the original LS method \citep{sussman1994level}, the zero-contour of level set function denotes the interface, and a re-initialization is performed in order to maintain the LS function as a distance function. Owing to the errors of advection and re-initialization, mass conservation is the main drawback in the original LS method. The coupled VOF and LS (CLSVOF) method \citep{sussman2000coupled, sun2010coupled} and  conservative LS (CLS) method \citep{olsson2005conservative, olsson2007conservative, waclawczyk2015consistent}, which use an alternative LS function, address this drawback.

Recently, the phase field (PF) method, another branch of interface capturing methods, has emerged as a promising method for multiphase flow simulations. It is based on two fundamental equations: Cahn-Hilliard (CH) and Allen-Cahn (AC) equations, dating back to the middle of the twentieth century  \citep{cahn1961spinodal,allen1979microscopic}. The CH equation can be conveniently reformulated into conservative form, making it a popular choice  \citep{jacqmin1999calculation,ding2007diffuse}, while the downside to the CH equation is that it needs to deal with fourth order spatial derivatives. On the other hand, although the original AC equation only requires second order spatial derivatives, it cannot conserve mass. Inspired by the conservative LS method \citep{olsson2005conservative} and an equivalent AC equation \citep{sun2007sharp}, the AC equation is reformulated into the conservative form incorporating the advantages of both the original CH and AC equations \citep{chiu2011conservative}. Recently, by restricting the free parameters of the AC equation, a finite difference scheme with central differencing was shown to guarantee the boundedness of the phase field  \citep{mirjalili2020conservative}.

We now turn our attention to exterior calculus (EC) and discretizations based on EC applied to fluid flow simulations. Smooth exterior calculus theory, as a powerful mathematical
tool, was  originally developed by the mathematician  Cartan  \citep{cartan1899certaines}. Exterior calculus deals with the calculus of differential forms, which can be thought of as ``the things which occur under integral signs" \citep{flanders1963differential}. For example, in three dimensional space, the integrations of a vector field over lines and surfaces  lead to 
the 1-form and 2-form: $\mathbf{u}\cdot d\mathbf{l}$ and $\mathbf{f}\cdot d \mathbf{A}$, respectively. Also, 3-form: $cdV$ appears in scalar volume integrations and 0-form can be considered as scalar functions. The other key quantities in exterior calculus include exterior derivative ($d$), Hodge star
operator ($*$) and exterior/wedge ($\wedge$) product. A good review on these quantities and differential forms is available in Ref.  \citep{perot2014differential}. Discrete exterior calculus (DEC) is the discretization of smooth exterior calculus on simplicial complexes and it is a bridge between smooth exterior calculus theory and computational science. Another discretization is finite element exterior calculus \citep{arnold2010finite}.  Generally, one approach approximating the smooth exterior calculus operators is to develop a method for the entire calculus by only using discrete geometrical and combinatorial operations  \citep{hirani2003discrete}.

Consider the key advantages of DEC: it is coordinate independent; works in arbitrary dimensions; has superior conservation
properties.  A series of works  \citep{elcott2007stable,mullen2009energy,hirani2015numerical,mohamed2016discrete,jagad2020investigation,jagad2021primitive,jagad2021effects} focusing on application of DEC to computational fluid dynamics appeared subsequent to the fundamental work by Hirani \citep{hirani2003discrete}. One of the early works was the development of a DEC scheme for incompressible, inviscid flow, which works on arbitrary simplicial meshes, and preserves discrete circulation and avoids numerical diffusion of vorticity \citep{elcott2007stable}. Later,
a simple, unconditionally stable, fully Eulerian energy-preserving DEC scheme, without numerical viscosity for incompressible and inviscid flow (Euler equation), was reported \citep{mullen2009energy}. A DEC scheme for Darcy flows was also developed by rewriting the governing equations in the standard vector differential notation to the differential form, and then discretizing them on a simplicial complex and its dual  \citep{hirani2015numerical}. Besides incompressible inviscid flows (Euler equation), DEC has also been extended to discretize the incompressible Navier-Stokes equations  \citep{mohamed2016discrete,jagad2020investigation,jagad2021primitive} for flows on arbitrary 2D surfaces. At present, the application of DEC has been limited to simulations of single phase fluid flow. In the present work, we develop physically compatible discretizations, based on exterior calculus formulations, of two-phase flows for simplicial meshes. We first develop the conservative phase field method in the EC framework in Section \ref{the_phase_field_method}. Then, we demonstrate the verification of the method for advection in Section \ref{verification}. In Section \ref{physical_phenomena}, we present simulation results for various physical phenomena governed by the full Navier-Stokes equations for two-phase immiscible fluid flow. Some conclusions are presented in Section \ref{conclusions}.

\section{Conservative phase field method in DEC framework \label{the_phase_field_method}}
\subsection{Governing equations in vector calculus notation}

Consider two immiscible fluids, which occupy an open bounded domain $\Omega$ with boundary $\partial\Omega$.  These two fluids separate the physical domain $\Omega$ into two subdomains $\Omega_{1}$ and $\Omega_{2}$ with boundaries $\partial\Omega_{1}$ and $\partial\Omega_{2}$, respectively, and $\Gamma=\partial\Omega_{1}\cap\partial\Omega_{2}$ is the interface between these two fluids. The governing equations comprising of the incompressible NS equations for each fluid in the subdomain $\Omega_{\alpha}$ $(\alpha = 1, 2)$ and the jump condition at the interface are as follows.
 \begin{gather}
\nabla\cdot\mathbf{u}_{\alpha}=0\label{sepreate_continuity_eqn},\\
\rho_{\alpha} \frac{\partial \mathbf{u}_{\alpha}}{\partial t} +\rho_{\alpha} (\mathbf{u}_{\alpha}\cdot \nabla )\mathbf{u}_{\alpha}=
-\nabla p_{\alpha} + \nabla\cdot(2\mu_{\alpha}\mathbf{D}_{\alpha})+
\rho_{\alpha} \mathbf{g}\label{seperate_momentum_eqn},\\
[p]_{\Gamma}\mathbf{n}=2[\mu\mathbf{D}]_{\Gamma}\mathbf{n}+\sigma\kappa\mathbf{n}\label{jump_condition}.
\end{gather}
where $\rho_{\alpha}$, $\mu_{\alpha}$, $\mathbf{u}_{\alpha}$, $p_{\alpha}$, $\mathbf{g}$ and $\mathbf{D}_{\alpha}=(\nabla\mathbf{u}_{\alpha}+(\nabla\mathbf{u}_{\alpha})^{T})/2$  are the density, the viscosity, the velocity field, the pressure field, the gravity and the rate of deformation tensor in each fluid, respectively. $\sigma$, $\kappa$ and $\mathbf{n}$ are the surface tension coefficient, curvature and unit normal vector on the interface, respectively. Here for arbitrary tensor $\mathbf{T}$, its jump at the interface is defined as $[\mathbf{T}]_{\Gamma}:=\mathbf{T}_{1}|_{\Gamma}-\mathbf{T}_{2}|_{\Gamma}$, when $\mathbf{n}$ points into subdomain $\Omega_{1}$, and vice versa. Equation (\ref{jump_condition}) is the jump condition at the interface, i.e., essentially an expression of continuity of traction (the balance among the pressure jump, surface tension force and the jump of normal viscous stress). Due to such fluid physical properties jump across the interface, the accurate interface location is key information in the two-phase flow simulation. We now introduce the notion of phase field variable in order to capture the interface $\Gamma$.

Based on the free energy of a nonuniform composition system, the phase-field (PF) model emerged as a promising approach for interface capturing in two-phase flow simulations \citep{jacqmin1999calculation}. In the aforementioned nonuniform composition system, the free energy $F$ is given by \citep{jacqmin1999calculation,mirjalili2021consistent}
 \begin{eqnarray}
F(\phi)=\frac{1}{2}\int_{\Omega}\epsilon^{2}|\nabla\phi|^{2}dV+\int_{\Omega}W(\phi)dV,
\label{Free_energy}
\end{eqnarray}
where function $\phi: [0,\infty] \times \Omega \to [-1, 1]$ is the  smooth PF function, $\phi = +1$ and $\phi=-1$ distinguish the pure phases occupying $\Omega_{1}$ and $\Omega_{2}$; $\epsilon$ is the interface thickness parameter and $W(\phi)=(1-\phi^{2})^{2}/4$ is the double-well potential. Two 
formulations of the fundamental governing equation of PF function exist: the CH equation and the AC equation. The change of phase field function in AC and CH equation are approximated as  being proportional to the free energy functional derivative and the diffusion of the functional derivative, respectively. One limitation of the original AC equation is its nonconservative form, although only second order spatial derivative appears in it, as compared to the CH equation.  After the precursor work of AC equation in curvature-driven flow \citep{sun2007sharp} and conservative level set method \citep{olsson2005conservative}, the reformulated conservative AC equation is derived \citep{chiu2011conservative}.

With the introduction of the PF function as the means of tracking the interface, the nondimensional incompressible NS equations in whole domain formulation \citep{scardovelli1999direct} coupled with conservative AC equation \citep{chiu2011conservative}  are as follows.
\begin{gather}
\nabla\cdot\mathbf{u}=0\label{continuity_eqn},\\
\rho \frac{\partial \mathbf{u}}{\partial t} +\rho (\mathbf{u}\cdot \nabla )\mathbf{u}=
-\nabla p + \frac{1}{Re}\nabla\cdot(2\mu\mathbf{D})+
\frac{1}{Bo}\kappa\mathbf{n}\delta(d)+\rho \mathbf{g}_{u}\label{momentum_eqn},\\
\frac{\partial\phi}{\partial t}+\nabla\cdot(\phi\mathbf{u})=
\nabla\cdot\left[
  \gamma\left(\epsilon\nabla\phi-  \phi(1-\phi)\frac{\nabla\phi}{|\nabla\phi|}\right)
\right]\label{pf_eqn},
\end{gather}
where $\mathbf{g}_{u}$ is the unit gravitational force, $\epsilon$ and $\mathbf{\gamma}$ are two positive free model parameters.  Two nondimensional parameters are: the Reynolds number defined as  $Re = L^{3/2}\sqrt{g}\rho_{c}/\mu_{c}$, and the Bond number defined as $Bo =
\rho_{c}gL^{2}/\sigma$.  Here, $L$ is the characteristic length of the physical phenomenon, $g$ is the gravitational acceleration,  $\rho_{c}$ is a characteristic (or reference) density and $\mu_{c}$ is the characteristic (or reference) viscosity. The density and viscosity of one of the phases are chosen as the characteristic quantities. The dimensionless density and viscosity of the two phase mixture are $\rho = \phi\rho_{1}/\rho_{c} + (1-\phi)\rho_{2}/\rho_{c}$ and $\mu =  \phi \mu_{1}/\mu_{c}+ (1-\phi)\mu_{2}/\mu_{c}$, respectively. Other quantities in the equations include $\delta$ the delta function and $d=|s|$ the distance function, where the signed distance function $s$ is defined as 
\begin{equation}
s(x)=\sgn(x)\min\limits_{y\in\Gamma, x\in\Omega}|x-y|, \qquad\qquad \sgn(x)=\left\{
\begin{array}{ccc}
1       &      & \forall x\in\Omega_{1}\\
-1     &      & \forall x\in\Omega_{2},
\end{array} \right.
\label{sign_distance}
\end{equation} 

In general, a kernel function \citep{sun2007sharp,chiu2011conservative}, which describe the normal variation of $\phi$ to the interface, is introduced for deriving conservative PF equation (\ref{pf_eqn}) as 
\begin{eqnarray}
\phi(x) = \frac{1}{2}\left [1+\tanh\left(\frac{s(x)}{2\epsilon}\right)\right].
\label{Kernel}
\end{eqnarray}

Therefore, the above unit normal $\mathbf{n}$ and the curvature $\kappa$ of the interface is mathematically expressed in terms of the PF function as follows.
\begin{equation}
\mathbf{n}=\frac{\nabla\phi}{|\nabla\phi|}, \qquad\qquad \kappa=\nabla\cdot \left(\frac{\nabla\phi}{|\nabla\phi|}\right ).
\end{equation}

\subsection{Exterior calculus notation}
In order to facilitate the discretization of two phase NS equations under the exterior calculus framework, we first express all relevant differential equations in exterior calculus notation. Differential forms are the key entities in the exterior calculus framework originally pioneered by \'{E}lie Cartan, and offer a coordinate independent approach to multivariable calculus in differential geometry and tensor calculus. In general, the scalars and vectors in differential equations operated upon by the differential operators: gradient $\nabla$, curl $\nabla\times$, divergence $\nabla\cdot$ and Laplace operator $\Delta$, can be transformed to equations involving \textit{differential forms} with exterior calculus operators: exterior derivative $d$, Hodge star $*$ and an exterior or wedge product $\wedge$. The flat operator ($\flat$) is an isomorphism that maps the vector fields to 1-forms, and is a bridge connecting vector differential equations and their exterior calculus versions.

If $\Omega$ is an open subset of $\mathbb{R}^3$ then one can use the  identity  $(\mathbf{u}\cdot \nabla )\mathbf{u}=\frac{1}{2}\nabla(\mathbf{u}\cdot\mathbf{u})-\mathbf{u}\times(\nabla\times\mathbf{u})$ in the momentum equation \eqref{momentum_eqn}  and use the fact that $(\mathbf{u}\times(\nabla\times\mathbf{u}))^{\flat} =
  \pm * (\mathbf{u}^{\flat}\wedge * d \mathbf{u}^{\flat})$. However, to develop a discretization that is valid on surfaces we will avoid the use of $\mathbf{u}\times(\nabla\times\mathbf{u})$. This is achieved by replacing $(\mathbf{u}\cdot \nabla) \mathbf{u}$ in~\eqref{momentum_eqn} by the equivalent, coordinate and dimension invariant form $\mathcal{L}_{\mathbf{u}} \mathbf{u}^\flat - \frac{1}2 d(\mathbf{u}\cdot \mathbf{u})$. See~\cite[p. 588]{AbMaRa1988}. Here $\mathcal{L}$ is the Lie derivative operator. Then using the Cartan formula~\cite[p. 588]{AbMaRa1988} we get
  \begin{align}
    \begin{split}
    ((\bu \cdot \nabla)\bu)^\flat = \mathcal{L}_\bu \bu^\flat - \frac{1}2 d(\bu \cdot \bu) =
    i_\bu d\bu^\flat + \frac{1}2 d (\bu \cdot \bu) &= (-1)^N *(\bu^\flat \wedge * d\bu^\flat) + \frac{1}2 d (\bu \cdot \bu)\\
    & = (-1)^N * (\bu ^\flat \wedge * d\bu^\flat) + (-1)^{N-1} d * (*\bu^\flat \wedge \bu^\flat)\, ,
    \label{convective_term}
  \end{split}
  \end{align}
  where $i_\bu d\bu^\flat$ is the 1-form defined by $i_\bu d\bu^\flat (\mathbf{v}) = d\bu^\flat(\bu,\mathbf{v})$ for all vector fields $\mathbf{v}$ on $\Omega$ and $N$ is the space dimension 2 or 3.

Applying the flat operator to terms of the two-phase NS momentum equation~\eqref{continuity_eqn} and using~\eqref{convective_term} the velocity divergence, convective term, pressure term and gravity term are transformed into exterior calculus notation involving 1-form $\mathbf{u}^{\flat}$ as 
\begin{eqnarray}
 \nabla\cdot\mathbf{u}&=&* d * \mathbf{u}^{\flat}\, ,\\
  \rho((\bu \cdot \nabla)\bu)^\flat &=&  
  (-1)^N \rho \wedge * (\bu ^\flat \wedge * d\bu^\flat) +
  (-1)^{N-1} \rho \wedge d * (*\bu^\flat \wedge \bu^\flat)\, ,\\
  (\nabla p)^{\flat}&=&d p\, ,\\
  (\rho \mathbf{g}_{u})^{\flat}&=&\rho \wedge \mathbf{g}_{u}^{\flat}\, ,
\end{eqnarray}
where $\mathbf{g}_{u}^{\flat}$ represents the unit gravitational force 1-form and $p$ is the  0-form pressure.

Due to the viscosity variation across the interface, the viscous stress term can be decomposed as 
\begin{eqnarray}
\nabla\cdot(2\mu\mathbf{D})=\mu\Delta\mathbf{u}+2\mathbf{D}\nabla\mu=\mu\Delta\mathbf{u}+2[\mu]_{\Gamma}\mathbf{D}\nabla\phi,
\end{eqnarray}
and $\mathbf{D}\nabla\phi$ can be rewritten in component form as 
\begin{eqnarray}
\mathbf{D}\nabla\phi=(\mathbf{D}\mathbf{n})\cdot\mathbf{n}\nabla\phi + \sum_{i=1}^{N-1}(\mathbf{D}\mathbf{n})\cdot\frac{\mathbf{t}_{i}}{|\mathbf{t}_{i}|}\mathbf{t}_{i}=D_{nn}\nabla\phi + \sum_{i=1}^{N-1}D_{\tilde{t_{i}}n}\mathbf{t}_{i},
\end{eqnarray}
where $\mathbf{t}_{i}$ is the tangent vector to the interface. Moreover, $\mathbf{t}_{i}\cdot\nabla\phi=0$, $\mathbf{t}_{i}\cdot\mathbf{t}_{j}=|\nabla\phi|^{2}\delta_{ij}$ (Kronecker delta) and $|\nabla\phi|=|\mathbf{t}_{i}|$. $D_{nn}=(\mathbf{D}\mathbf{n})\cdot\mathbf{n}=\frac{\partial u_{n}} {\partial n}$ and $D_{\tilde{t_{i}}n}=(\mathbf{D}\mathbf{n})\cdot\frac{\mathbf{t}_{i}}{|\mathbf{t}_{i}|}=\frac{1}{2}(\frac{\partial u_{n}}{\partial \tilde{t_{i}}}+\frac{\partial u_{\tilde{t_{i}}}}{\partial n})$ are the normal and shear strain rate of the interface, where $u_{n}$ and $u_{\tilde{t_{i}}}$is the normal and tangent speed on the interface. 
The normal speed $u_{n}$ and tangent speed $u_{\tilde{t_{i}}}$ in exterior calculus notation are given by
\begin{eqnarray}
u_{n}=\frac{(-1)^{N-1}*[(*\mathbf{u}^{\flat})\wedge d \phi]}{\{(-1)^{N-1}*[(*d\phi)\wedge d\phi]\}^{1/2}},\\
u_{\tilde{t_{i}}}=\frac{(-1)^{N-1}*[(*\mathbf{u}^{\flat})\wedge \mathbf{t}_{i}^{\flat}]}{\{(-1)^{N-1}*[(*d\phi)\wedge d\phi]\}^{1/2}}.
\end{eqnarray}
Applying the flat operator on the viscous stress term and {performing} the notation transformation, we obtain
\begin{gather}
\left(\frac{1}{Re}\nabla\cdot(2\mu\mathbf{D})\right)^{\flat}=(-1)^{N+2}\frac{\mu}{Re}*d*d\mathbf{u}^{\flat}+\frac{2[\mu]_{\Gamma}}{Re}\left[\left(D_{nn}\nabla\phi\right)^{\flat}  +\sum_{i=1}^{N-1}\left(D_{\tilde{t_{i}}n}\mathbf{t}_{i}\right)^{\flat}\right]\label{deformation_tensor_EC},
\end{gather}
where the normal strain rate $D_{nn}$ and the shear strain rate $D_{\tilde{t_{i}}n}$ under exterior calculus notation are
\begin{gather}
D_{nn}=\frac{(-1)^{N-1}*[(*d\phi)\wedge d u_{n}]}{(-1)^{N-1}*[(*d\phi)\wedge d\phi]\}^{1/2}}\label{normal_stress_jump_EC},\\
D_{\tilde{t_{i}}n}=\frac{1}{2}\left[\frac{(-1)^{N-1}*[(*d\phi)\wedge d u_{\tilde{t_{i}}}]}{\{(-1)^{N-1}*[(*d\phi)\wedge d\phi]\}^{1/2}} + \frac{(-1)^{N-1}*[(*\mathbf{t}_{i}^{\flat})\wedge d u_{n}]}{\{(-1)^{N-1}*[(*d\phi)\wedge d\phi]\}^{1/2}} \right]
\label{shear_stress_jump_EC}.
\end{gather}

The CSF model \citep{brackbill1992continuum} offers an alternative approach to represent the surface tension force $F_{s}=\sigma\kappa\mathbf{n}\delta(d)$ under a volume force form: $F_{v}=\sigma\kappa\nabla\phi$. Therefore, the surface tension term can be expressed under volume force form and in the exterior calculus notation
\begin{eqnarray}
[\frac{1}{Bo}\kappa\mathbf{n}\delta(d)]^{\flat}=\frac{1}{Bo}(\kappa\nabla\phi)^{\flat},
\end{eqnarray}
where the curvature in exterior calculus notation is expressed as
\begin{eqnarray}
\kappa=*d*\left(\frac{\nabla\phi}{|\nabla\phi|}\right)^{\flat}.
\end{eqnarray}

 All terms of conservative PF equation (\ref{pf_eqn}) can be denoted under exterior calculus notation by considering $\phi$ as a 0-form:
\begin{eqnarray}
\nabla\cdot(\phi\mathbf{u})&=&*d*(\phi\wedge\mathbf{u}^{\flat})\, ,\\
\Delta\phi&=&*d*d\phi\, ,\\
\nabla\cdot\left(\phi(1-\phi)\frac{\nabla\phi}{|\nabla\phi|}\right)&=&*d*\left(\phi(1-\phi)\frac{\nabla\phi}{|\nabla\phi|}\right)^{\flat}.
\end{eqnarray}
Therefore, the entire set of the dimensionless form of the governing equations 
(\ref{continuity_eqn})-(\ref{pf_eqn})
in exterior calculus notation are:
\begin{gather}
*d*\mathbf{u}^{\flat}=0\label{continuity_eqn_EC},\\
\begin{split}
\rho\wedge \frac{\partial \mathbf{u}^{\flat}}{\partial t}
+(-1)^{N} \rho \wedge [* (\mathbf{u}^{\flat}\wedge * d \mathbf{u}^{\flat})]=-dp
+(-1)^{N}\frac{\mu}{Re}*d*d\mathbf{u}^{\flat}
+\frac{2[\mu]_{\Gamma}}{Re}\left[\left(D_{n}\nabla\phi\right)^{\flat}+\sum_{i=1}^{N-1}\left(D_{\tilde{t_{i}}n}\mathbf{t}_{i}\right)^{\flat}\right]\\
+\frac{1}{Bo}\left(\kappa\nabla\phi\right)^{\flat}
+\rho\wedge\mathbf{g}_{u}^{\flat}
+\frac{(-1)^{N}}{2}\rho \wedge d\{*[(*\mathbf{u}^{\flat}) \wedge \mathbf{u}^{\flat}]\}\label{momentum_eqn_EC},\\
\end{split}\\
\frac{\partial\phi}{\partial t}+*d*(\phi\wedge\mathbf{u}^{\flat})=\gamma*d*\left(\epsilon d\phi - \left(\phi(1-\phi)\frac{\nabla\phi}{|\nabla\phi|}\right)^{\flat}\right)\label{PF_eqn_EC}.
\end{gather}


\subsection{Spatial discretization}

\begin{figure}[htbp]
\centering
\includegraphics[width=0.5\linewidth]{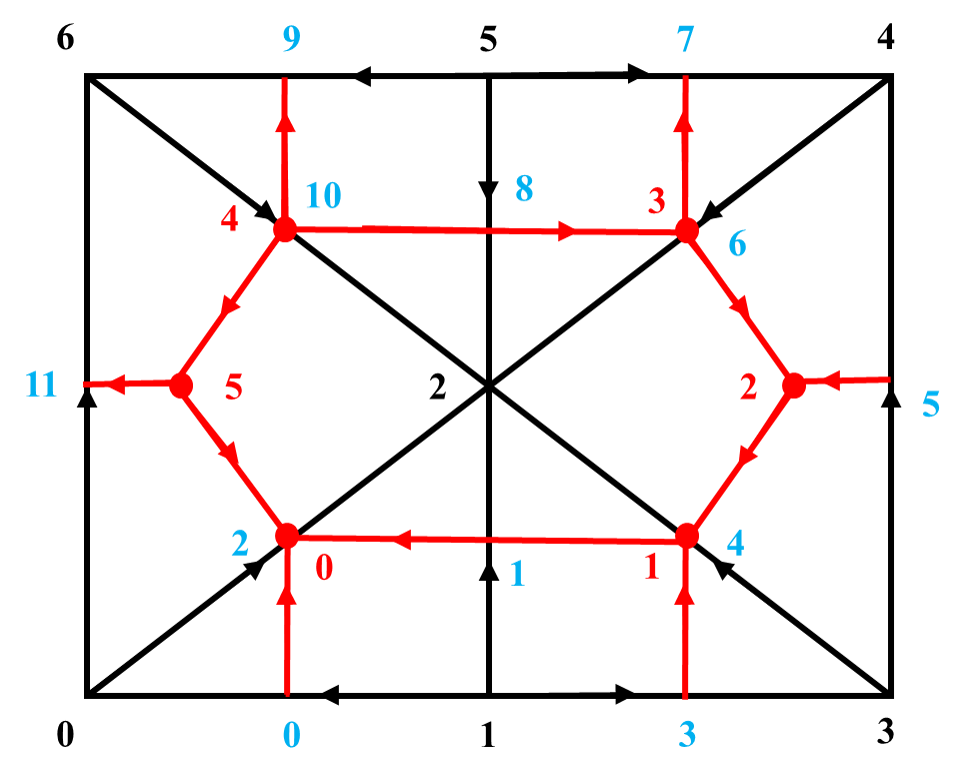}
\caption{A sample 2D mesh comprising of primal simplices (in black color) and dual simplices (in red color). The orientation of primal $1-simplex$ can be arbitrary and induce the orientation of its dual $1-cell$ by rotating $90$ degrees counterclockwise. The orientation of primal $2-simplex$ and dual $2-cell$ is counterclockwise.}
\label{mesh}
\end{figure}

 The $N$-dimensional physical space is discretized with simplicial complex $K$. A $k$-dimensional simplex can be denoted by its vertices as $\sigma^{k}=[v_{0},v_{1},...,v_{k-1},v_{k}]$, where the subscripts represent the vertex indices and the order of the vertices defines the orientation of the simplex. The geometrical entities comprising low dimensional simplices is intuitive: a 0-simplex is a vertex, a 1-simplex is a line segment, a 2-simplex is a triangle and a 3-simplex is a tetrahedron. The notation $\sigma^{k}\in K$ represents that $k$-dimensional simplex is in the simplicial complex $K$. The dual complex $\star K$ is the duality of simplicial complex $K$. For a $k$-dimensional simplex $\sigma^{k}\in K$, its dual $\star\sigma^{k}\in \star K$ is an $(N-k)$-cell. In a two dimensional simplicial complex, the orientation of the primal 2-simplices (triangles) and the dual 2-cells (polygons) is assumed to be counterclockwise, while that of the primal 1-simplices (edges) can be arbitrary. 
The orientation of dual 1-cells (edges) is obtained by rotating the primal 1-simplices $90$ degrees counterclockwise. Presently, we consider only the Delaunay simplicial mesh and the circumcentric dual. These choices are not inherent limitations and in fact, the methods considered presently are generally applicable to non-Delaunay meshes and other choices of the dual. However, to limit the scope of the present work, these will not be considered.  Figure \ref{mesh} illustrates a simple 2D mesh. Let $N_{k}$ be the number of $k$-simplices, then $N_{0}=7$, $N_{1}=12$ and $N_{2}=6$ for the example mesh. In this paper, the general Delaunay triangulation and the triangulation, in which every  dual 2-cell degenerates to a rectangle, are referred to as irregular triangulation and regular triangulation, respectively.

\begin{figure}[htbp]
\centering
\includegraphics[width=0.5\linewidth]{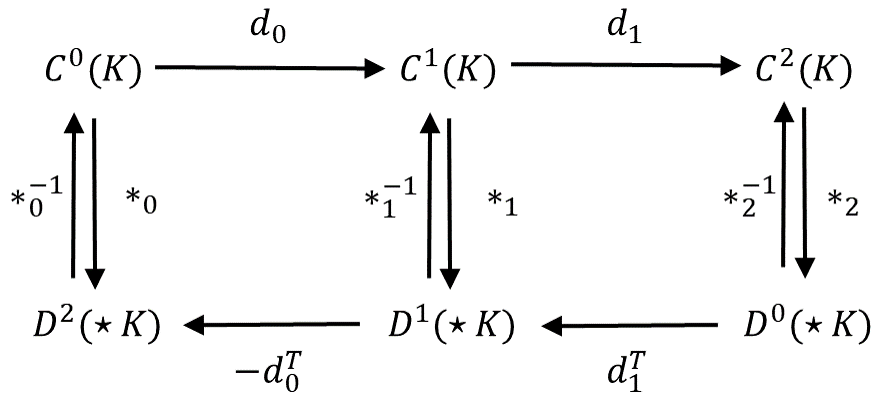}
\caption{The DEC operators diagram of discrete differential form (cochain) on 2D primal and dual meshes.}
\label{operator}
\end{figure}

 The discrete k-forms can be considered essentially as scalars obtained by integrating smooth k-forms over k-simplexes, and the space of discrete k-forms defined on primal and dual mesh complexes are denoted by $C^{k}(K)$ and $D^{k}(\star K)$, respectively. For example, for smooth velocity 1-form $\mathbf{u}\cdot d \mathbf{l}$, the discrete velocity 1-form can be defined on primal edge $\sigma^{1}$ as: $v=\int_{\sigma^{1}}\mathbf{u}\cdot d \mathbf{l}\in C^{1}(K)$ or on dual edge $*\sigma^{1}$: $u=\int_{*\sigma^{1}}\mathbf{u}\cdot d \mathbf{l}\in D^{1}(\star K)$. The mapping between these spaces, $C^{k}(K)$ and $D^{k}(\star K)$, with the discrete Hodge star and exterior derivative operators is shown in the cochain diagram in Figure \ref{operator}.

The discrete exterior derivative operator $d_{k}$ maps a primal k-form to a primal $(k+1)$-form, which is a sparse matrix defined as the transpose of the boundary operator for the $(k+1)$-simplices. With the negative sign for the $d^{T}_{0}$ operator in 2D, the transpose of the $d_{(N-k-1)}$ is the discrete derivative operator for dual k-form.

The discrete Hodge star $*_{k}$ maps a primal k-form to dual (N-k)-form and $*^{-1}_{k}$ is the inverse map of $*_{k}$. The $*_{k}$ is a diagonal matrix, owing to the choice of a circumcentric dual, whose diagonal element is the ratio between the volume of the dual $(N-k)$-cell $\star \sigma^{k}$ and the primal $k$-simplex $\sigma^{k}$, i.e., $|\star\sigma^{k}|/|\sigma^{k}|$.

\subsection{Boundedness analysis  \label{section_boundedness_analysis}}
In two-phase flow simulations, two desirable features are avoiding interface diffusion and keeping the boundedness of density and viscosity. In conservative PF equation, the compressive term $\epsilon\nabla\cdot[\phi(1-\phi)\frac{\nabla\phi}{|\nabla\phi|}]$  is conducive towards a sharp interface because this term counteracts numerical diffusion near the interface. We recall that $\phi$ lives on the dual nodes. In order to keep the boundedness of density and viscosity,  eliminating the numerical oscillations near the interface is desirable and some previous work in two phase flow simulation use a technique called redistribution algorithm to achieve it \citep{harvie2000new,chiu2011conservative}. Because of the linear relationship between viscosity and density with the PF function $\phi$, it is evident that the boundedness of density and viscosity is equivalent to the boundedness of PF function $\phi$ for PF method. Recently, a central finite different explicit Euler forward schemes for conservation PF equation with appropriate parameters $\epsilon$ and $\gamma$ has been proved to guarantee bounded $\phi$ \citep{mirjalili2020conservative}. 
  
We now discuss the boundedness analysis in DEC framework and further extend the proof of the boundedness for explicit Euler forward scheme to predictor-corrector method. The principle of mathematical induction is employed for the proof. Initially, the $\phi$ boundedness statement holds, i.e. $\phi\in [0,1]$. Then, we assume the $\phi$ boundedness statement holds for an arbitrary time step $n$, and prove the $\phi$ boundedness statement in time step $n+1$.

All terms of the conservative PF equation in dual 0-cell $\star \sigma^{2}$ can be expressed in general local DEC notation straightforwardly as
\begin{gather}
  \nabla\cdot(\mathbf{u}\phi)(\star \sigma^{2})=
  \left<*d*(\phi\wedge\mathbf{u}^{\flat}),\star\sigma^2\right>=
  \frac{-|\star\sigma^2|}{s(\star\tilde{\sigma}^{1})|\sigma^{2}|}
  \sum_{\sigma^{1}\prec\sigma^{2}}\sgn(\partial\sigma^{2},\sigma^{1})\frac{|\sigma^{1}|}{|\star\sigma^{1}|}
  \sum_{\star\hat{\sigma}^{2}\prec\star\sigma^{1}}\frac{\phi(\star\hat{\sigma}^{2})}{2}
  \left<\mathbf{u}^{\flat},\star\sigma^{1}\right>,\\
  \Delta\phi(\star\sigma^2)=
  \left<*d*d\phi,\star\sigma^2\right>=
  \frac{-|\star\sigma^2|}{s(\star\tilde{\sigma}^{1})|\sigma^{2}|}
  \sum_{\sigma^{1}\prec\sigma^{2}}\sgn(\partial\sigma^{2},\sigma^{1})\frac{|\sigma^{1}|}{|\star\sigma^{1}|}
  \sum_{\star\hat{\sigma}^{2}\prec\star\sigma^{1}}\sgn(\partial\star\sigma^{1},\star\hat{\sigma}^{2})\phi(\star\hat{\sigma}^{2}),\\
  \nabla\cdot(\phi(1-\phi)\frac{\nabla\phi}{|\nabla\phi|})(\star\sigma^2)=
  -\frac{|\star\sigma^2|}{s(\star\tilde{\sigma}^{1})|\sigma^{2}|}
  \sum_{\sigma^{1}\prec\sigma^{2}}\sgn(\partial\sigma^{2},\sigma^{1})\frac{|\sigma^{1}|}{|\star\sigma^{1}|}
  \left<\left(\phi(1-\phi)\frac{\nabla\phi}{|\nabla\phi|}\right)^{\flat},\star\sigma^{1}\right>,
\end{gather}
where $s(\star\tilde{\sigma}^{1})=\pm1$ and the following rule decides the sign of $s(\tilde{\sigma}^{1})$. Assign $\tilde{\sigma}^{1}$ the orientation induced form $\sigma^{2}$ and if $\star\tilde{\sigma}^{1}$ points away from $\star\sigma^2$, then $s(\star\tilde{\sigma}^{1})=-1$, otherwise, $s(\star\tilde{\sigma}^{1})=1$. Presently, the orientation of primal 2-simplices and dual 2-cells are assumed to be counterclockwise, so that here $s(\star\tilde{\sigma}^{1})=1$. Moreover, $\sgn(\partial\sigma^{2},\sigma^{1})$ and $\sgn(\partial\star\sigma^{1},\star\hat{\sigma}^{2})$ denote the relative orientation between the boundary of a primal simplex/dual cell and its top dimensional proper face. For the compressive flux dual 1-form, we define its discrete form as
\begin{eqnarray}
\left<\left(\phi(1-\phi)\frac{\nabla\phi}{|\nabla\phi|}\right)^{\flat},\star\sigma^{1}\right>:=\sum_{\star\hat{\sigma}^{2}\prec\star\sigma^{1}}\left(\frac{1}{2}\phi(1-\phi)\frac{\nabla\phi}{|\nabla\phi|}\right)(\star\hat{\sigma}^{2})\cdot(\star\vec{\sigma}^{1}).
\label{discret_compressive_flux_1_form}
\end{eqnarray}
Given a $p$-chain $c$ and a discrete p-form $\alpha$, the discrete Stokes' theorem $\left<d\alpha,c\right>=\left<\alpha,\partial c\right>$ is independent of the choice of primal simplices/dual cells orientation. Therefore, we can always set $\sgn(\partial\sigma^{2},\sigma^{1})=1$ (the orientation of $\sigma^{1}$ is induced by $\sigma^{2}$) without losing generalization, which leads to $\sgn(\partial\star\sigma^{1},\star\sigma^{2})=1$ and $\sgn(\partial\star\sigma^{1},\star\hat{\sigma}^{2})=-1$ for $\star\hat{\sigma}^{2}\neq\star\sigma^{2}$. Therefore, the local DEC expression of conservative PF equation can be written under explicit Euler forward scheme as
\begin{gather}
\begin{split}
  \frac{\phi^{n+1}(\star\sigma^{2}) - \phi^{n}(\star\sigma^{2})}{\Delta t} =
  -\frac{|\star\sigma^{2}|}{|\sigma^{2}|}\sum_{\sigma^{1}\prec\sigma^{2}}\frac{|\sigma^{1}|}{|\star\sigma^{1}|}
  \sum_{\star\hat{\sigma}^{2}\prec\star\sigma^{1}}
  \left( \epsilon\gamma \sgn(\partial\star\sigma^{1},\star\hat{\sigma}^{2})\phi^{n}(\star\hat{\sigma}^{2})\right)\\
  +\frac{|\star\sigma^{2}|}{|\sigma^{2}|}
  \sum_{\sigma^{1}\prec\sigma^{2}}\frac{|\sigma^{1}|}{|\star\sigma^{1}|}
  \sum_{\star\hat{\sigma}^{2}\prec\star\sigma^{1}}
  \left(\frac{\phi^{n}(\star\hat{\sigma}^{2})}{2}\left<\mathbf{u}^{\flat},\star\sigma^{1}\right>\right)\\
  +\gamma\frac{|\star\sigma^{2}|}{|\sigma^{2}|}
  \sum_{\sigma^{1}\prec\sigma^{2}}\frac{|\sigma^{1}|}{|\star\sigma^{1}|}
  \left<\left(\phi^{n}(1-\phi^{n})\frac{\nabla\phi^{n}}{|\nabla\phi^{n}|}\right)^{\flat},\star\sigma^{1}\right>.
\end{split}
\end{gather}
Combining with discrete continuity equation in general local DEC notation:
\begin{gather}
-\frac{|\star\sigma^{2}|}{|\sigma^{2}|}\sum_{\sigma^{1}\prec\sigma^{2}}\frac{|\sigma^{1}|}{|\star\sigma^{1}|}\left<\mathbf{u}^{\flat},\star\sigma^{1}\right>=0.
\end{gather}
The discrete PF equation can be reformulated to
\begin{multline}
  \phi^{n+1}(\star\sigma^{2}) =
  \left(1-\Delta t\gamma\epsilon\frac{|\star\sigma^{2}|}{|\sigma^{2}|}
    \sum_{\sigma^{1}\prec\sigma^{2}}\frac{|\sigma^{1}|}{|\star\sigma^{1}|} +
    \frac{1}{2}\Delta t \gamma\frac{|\star\sigma^{2}|}{|\sigma^{2}|}
    \sum_{\sigma^{1}\prec\sigma^{2}}\frac{|\sigma^{1}|}{|\star\sigma^{1}|}
    (1-\phi^{n})\mathbf{n}^{n}\cdot(\star\vec{\sigma}^{1}) \right)
  \phi^{n}(\star\sigma^{2})\\
  + \frac{|\star\sigma^{2}|}{|\sigma^{2}|}
  \sum_{\sigma^{1}\prec\sigma^{2}}\frac{|\sigma^{1}|}{|\star\sigma^{1}|}
  \left( \frac{1}{2}\Delta t\left<\mathbf{u}^{\flat},\star\sigma^{1}\right> +
    \Delta t\gamma\epsilon +
    \frac{1}{2}\Delta t \gamma(1-\phi^{n})\mathbf{n}^{n}\cdot(*\vec{\sigma}^{1}) \right)
  \phi^{n}(\star\tilde{\sigma}^{2}),
  \label{PF_eqn_local_DEC}
\end{multline}
where $\mathbf{n} = \nabla\phi/|\nabla\phi|$ and $\star\tilde{\sigma}^{2}\in\{\star\hat{\sigma}^{2}|\star\hat{\sigma}^{2}\prec\star\sigma^{1},\star\hat{\sigma}^{2}\neq\star\sigma^{2}\}$. Inspired by the proof in differential scheme \citep{mirjalili2020conservative}, we introduce auxiliary variables $\tilde{\phi}=1-\phi$ and $\tilde{\mathbf{n}} = \nabla\tilde{\phi}/|\nabla\tilde{\phi}|=-\mathbf{n}$, and substitute into equation (\ref{PF_eqn_local_DEC}). Hence, we obtain the equation of $\tilde{\phi}$, which is identical to that of $\phi$ and written as follows.
\begin{multline}
  \tilde{\phi}^{n+1}(\star\sigma^{2}) =
  \left(1-\Delta t\gamma\epsilon\frac{|\star\sigma^{2}|}{|\sigma^{2}|}
    \sum_{\sigma^{1}\prec\sigma^{2}}\frac{|\sigma^{1}|}{|\star\sigma^{1}|} +
    \frac{1}{2}\Delta t \gamma\frac{|\star\sigma^{2}|}{|\sigma^{2}|}
    \sum_{\sigma^{1}\prec\sigma^{2}}\frac{|\sigma^{1}|}{|\star\sigma^{1}|}
    (1-\tilde{\phi}^{n})\tilde{\mathbf{n}}^{n}\cdot(\star\vec{\sigma}^{1}) \right)
  \tilde{\phi}^{n}(\star\sigma^{2})\\
  + \frac{|\star\sigma^{2}|}{|\sigma^{2}|}
  \sum_{\sigma^{1}\prec\sigma^{2}}\frac{|\sigma^{1}|}{|\star\sigma^{1}|}
  \left( \frac{1}{2}\Delta t\left<\mathbf{u}^{\flat},\star\sigma^{1}\right> +
    \Delta t\gamma\epsilon +
    \frac{1}{2}\Delta t \gamma(1-\tilde{\phi}^{n})\tilde{\mathbf{n}}^{n}\cdot(*\vec{\sigma}^{1}) \right)
  \tilde{\phi}^{n}(\star\tilde{\sigma}^{2}).
\label{auxiliary_PF_eqn_local_DEC}
\end{multline}
It implies that if a choice of values for parameters $\gamma$ and $\epsilon$ assures positive $\phi$  in the next time step, the boundedness of $\phi$ is guaranteed. In order to keep $\phi$ positive in next time step, we have two inequalities:
\begin{gather}
  1-\frac{\Delta t\gamma|\star\sigma^{2}|}{|\sigma^{2}|}
  \sum_{\sigma^{1}\prec\sigma^{2}}\frac{|\sigma^{1}|}{|\star\sigma^{1}|}
  \left(\epsilon - \frac{1}{2}(1-\phi^{n})\mathbf{n}^{n}\cdot(\star\vec{\sigma}^{1}) \right)\geq 1 -
  \frac{\Delta t\gamma}{|\sigma^{2}|_{min}}
  \sum_{\sigma^{1}\prec\sigma^{2}}\left[\epsilon\left(\frac{|\sigma^{1}|}{|\star\sigma^{1}|}\right)_{max}
 + \frac{|\sigma^{1}|_{max}}{2}\right]
 \geq 0,\\
\frac{1}{2}\left\langle\mathbf{u}^{\flat},\star\sigma^{1}\right\rangle + \gamma\epsilon + \frac{1}{2} \gamma(1-\phi^{n})\mathbf{n}^{n}\cdot(\star\vec{\sigma}^{1})\geq 0, 
\end{gather}
and obtain the constrains on $\Delta t$, $\epsilon$ and $\gamma$ for guaranteeing boundedness of $\phi$:
\begin{gather}
\Delta t \leq \frac{|\sigma^{2}|_{min}}{3\gamma\left[\epsilon\left(\frac{|\sigma^{1}|}{|\star\sigma^{1}|}\right)_{max}
 + \frac{1}{2}|\sigma^{1}|_{max}\right]}\\
 \epsilon \geq \frac{1}{2}\left(
 \frac{\left|\left<\mathbf{u}^{\flat},\star\sigma^{1}\right>\right|_{max}}{\gamma}+|\star\sigma^{1}|_{max}\right)
\end{gather}

Hitherto, we proved that the first order explicit Euler forward scheme in the DEC framework can guarantee boundedness of $\phi$ by choosing proper values of the parameters. Now we extend the proof of $\phi$ boundedness to a higher order explicit time integration scheme, namely predictor-corrector scheme, which has second order accuracy. The predictor-corrector scheme consists of two steps, a predicted step followed by a corrector step. In the predictor step, we update $\phi$ from the values of $\phi$ at the previous time step ($\phi^{n}$), and denote it as $\phi_{n+1}^{*}$. In the corrector step, we update $\phi$ further from $\phi^{n}$ and $\phi_{n+1}^{*}$ and regard it as the current value ( $\phi^{n+1}$). The expressions of predictor and corrector steps are:
\begin{gather}
\phi_{n+1}^{*}=\phi^{n}+\Delta t f(t_{n},\phi^{n}),\\
\phi^{n+1} = \phi^{n} + \frac{\Delta t}{2}[f(t_{n},\phi^{n})+f(t_{n+1},\phi_{n+1}^{*})].
\end{gather}
Let $\phi_{E}^{n+1}=\phi_{n+1}^{*}$, where the subscript $E$ denotes explicit Euler forward scheme, which is used in the predicted step, so that the corrected step can be reformulated as
\begin{multline}
\phi^{n+1} = \phi^{n} + \frac{\Delta t}{2}[f(t_{n},\phi^{n})+f(t_{n+1},\phi_{E}^{n+1})]=\frac{1}{2}[2\phi^{n}+\Delta tf(t_{n},\phi^{n})+\Delta tf(t_{n+1},\phi_{E}^{n+1})]\\
=\frac{1}{2}[\phi^{n}+\phi_{E}^{n+1}+\Delta t f(t_{n+1},\phi_{E}^{n+1})]=\frac{1}{2}[\phi^{n}+\phi_{E}^{n+2}].
\end{multline}
Because $0\leq\phi^{n}\leq1$ and $0\leq\phi_{E}^{n+2}\leq1$, it is straightforward that $0\leq\phi^{n+1}\leq1$ in the predictor-corrector scheme. Thus,  the constraint on the parameters $\Delta t$, $\gamma$ and $\epsilon$ is the same  as that in explicit Euler forward scheme. In this paper, $\gamma$ and $\epsilon$ are set as $\frac{\left|\left<\mathbf{u}^{\flat},\star\sigma^{1}\right>\right|_{max}}  {|\star\sigma^{1}|_{max}}$ and $|\star\sigma^{1}|_{max}$, respectively, for all simulations.

\subsection{DEC notation}
To obtain the discrete exterior calculus form of the dimensionless PF description of two-phase NS equations(\ref{continuity_eqn_EC})-(\ref{PF_eqn_EC}), we first define discrete variables on their corresponding mesh objects. Then, the smooth differential forms and differential operators are replaced by their corresponding discrete counterparts. Presently, all of the discrete velocity 1-form $\mathbf{u}^{\flat}$ in the equations are  defined on the dual  edges, except one of the discrete velocity 1-form in the convection term of the momentum equation (\ref{momentum_eqn_EC}), i.e. the term $(-1)^{N+2}\rho\wedge[*(\mathbf{u}^{\flat}\wedge * d \mathbf{u}^{\flat} )]$ which is defined on dual edges. In this nonlinear product, it is natural to define one of its discrete  velocity 1-form $\mathbf{u}^{\flat}$ on dual edge and the other on the primal edge. Here, we define the first discrete velocity 1-form on the primal edges and the second on the dual edges. Consistently, the discrete density 0-form $\rho$, the discrete viscosity 0-form $\mu$, the discrete pressure 0-form $p$ and the discrete PF function 0-form $\phi$ are defined on dual points. 

The discretization of the smooth exterior calculus continuity equation (\ref{continuity_eqn_EC}) can be derived in matrix form directly by substituting appropriate discrete operators. Hence, we have
\begin{eqnarray}
*_{2}d_{1}*_{1}^{-1} U = 0,
\end{eqnarray}
where $U$ is the vector containing the discrete dual velocity 1-form $u$ for each dual edges. The DEC notation of momentum equation (\ref{momentum_eqn_EC}) with $N=2$ for two dimensional space is then expressed in matrix form as
\begin{multline}
W_{\rho}\frac{\partial U}{\partial t}  + W_{\rho}*_{1}W_{v}*_{0}^{-1}\left[ [-d_{0}^{T}]U + d_{b}V \right]= - d_{1}^{T}P + \frac{1}{Re}W_{\mu}*_{1}d_{0}*_{0}^{-1}\left[ [-d_{0}^{T}]U + d_{b}V \right] + W_{\rho} G_{u}\\ + \frac{1}{Bo}S
+\frac{2[\mu]_{\Gamma}}{Re}\left[N+T\right]-\frac{1}{2}W_{\rho}d_{1}^{T}K ,
\end{multline}
where $P$, $V$, $\Phi$, $G_{u}$ and $K$ are the vectors containing the dual pressure 0-form $p$, primal velocity 1-form $v$, PF function 0-form $\phi$, dual unit gravity 1-form $g_{u}$ and dual kinetic energy 0-form $k$, respectively. The diagonal matrices $W_{\rho}$, $W_{v}$ and $W_{\mu}$ contain the density 0-form $\rho$, tangential velocity 1-form $v$ and viscosity 0-form $\mu$, respectively, and represent the discrete wedge product. The discrete operator $[-d_{0}^{T}U]$ implies the discrete Stokes theorem on discrete velocity dual 1-form $u$ and $d_{b}$ is the boundary operator complement for the primal edge boundary of dual $2-cells$ which coincides with the domain boundary. $S$, $N$ and $T$ are the vectors containing discrete surface tension force dual 1-form, normal strain rate dual 1-form and shear strain rate dual 1-form, which are defined as
\begin{gather}
\left<\left(\kappa\nabla\phi\right)^{\flat},\star\sigma^{1}\right>:=\sum_{\star\sigma^{2}\prec\star\sigma^{1}}\left(\kappa\nabla\phi\right)(\star\sigma^{2})\cdot(\star\vec{\sigma}^{1}),\\
\left<\left(D_{nn}\nabla\phi\right)^{\flat},\star\sigma^{1}\right>:=\sum_{\star\sigma^{2}\prec\star\sigma^{1}}\left(D_{nn}\nabla\phi\right)(\star\sigma^{2})\cdot(\star\vec{\sigma}^{1}),\\
\left<\left(D_{\tilde{t_{1}}n}\mathbf{t}_{1}\right)^{\flat},\star\sigma^{1}\right>:=\sum_{\star\sigma^{2}\prec\star\sigma^{1}}\left(D_{\tilde{t_{1}}n}\mathbf{t}_{1}\right)(\star\sigma^{2})\cdot(\star\vec{\sigma}^{1}).
\end{gather}
where $\star\vec{\sigma}^{1}$ denotes the vector corresponding to $\star\sigma^{1}$ with the same direction as the orientation of $\star\sigma^{1}$.
The discrete $D_{nn}$ and $D_{\tilde{t_{1}}n}$ are obtained by discretizing equation (\ref{normal_stress_jump_EC}, \ref{shear_stress_jump_EC}) and the discrete primal-dual wedge product is implemented on top-dimensional 
simplex (primal/dual 2 cell in 2D). The definition of primal-dual wedge product on support volumes is discussed in reference  \citep{hirani2003discrete} and here we extend the definition on top-dimensional simplex (2D):
\begin{eqnarray}
\left<\alpha^{1}\wedge\beta^{1},\sigma^{2}\right>=\sum_{\sigma^{1}\prec\sigma^{2}}\frac{|V_{\sigma^{1}}\bigcap\sigma^{2}|}{|V_{\sigma^{1}}|}\left<\alpha^{1},\sigma^{1}\right>\left<\beta^{1},\star\sigma^{1}\right>,
\end{eqnarray}
where $\alpha^{1}$ and $\beta^{1}$ are 1-form. Also, we can write conservative PF equation under DEC notation in matrix form as
\begin{eqnarray}
\frac{\partial \Phi}{\partial t} + *_{2}d_{1}*_{1}^{-1}W_{u}\Phi = \epsilon\gamma*_{2}d_{1}*_{1}^{-1}d_{1}^{T}\Phi - \gamma*_{2}d_{1}*_{1}^{-1}F,
\end{eqnarray}
where the matrix $W_{u}$ represents the discrete wedge product and contains the dual velocity 1-form $u$, and $F$ is the vector containing discrete compressive flux dual 1-form as defined in equation (\ref{discret_compressive_flux_1_form}).


Using the predictor-corrector scheme and the midpoint integration to update PF function $\phi$ and dual velocity 1-form $u$, respectively, we can write a second order discrete PF description of two phase NS equation as
\begin{gather}
\frac{\Phi^{n+1*} - \Phi^{n}}{\Delta t} + *_{2}d_{1}*_{1}^{-1}W_{u}^{n}\Phi^{n} = \epsilon\gamma*_{2}d_{1}*_{1}^{-1}d_{1}^{T}\Phi^{n} - \gamma*_{2}d_{1}*_{1}^{-1}F^{n},\\
\frac{\Phi^{n+1} - \Phi^{n}}{\Delta t} + \frac{*_{2}d_{1}*_{1}^{-1}W_{u}^{n}(\Phi^{n}+\Phi^{n+1*})}{2} = \frac{\epsilon\gamma*_{2}d_{1}*_{1}^{-1}d_{1}^{T}(\Phi^{n}+\Phi^{n+1*})}{2} - \frac{\gamma*_{2}d_{1}*_{1}^{-1}(F^{n}+F^{n+1*})}{2},\\
\begin{split}
\frac{W_{\rho}^{n+1}(U^{n+1} - U^{n})}{\Delta t} + \frac{1}{2}W_{\rho}^{n+1}*_{1}W_{v}^{n}*_{0}^{-1}\left(\left[ [-d_{0}^{T}]U^{n+1} + d_{b}V^{n} \right]+\left[ [-d_{0}^{T}]U^{n} + d_{b}V^{n} \right]\right)=- d_{1}^{T}(P)^{n+1/2}\\
\frac{1}{2Re}W_{\mu}^{n+1}*_{1}d_{0}*_{0}^{-1}\left(\left[ [-d_{0}^{T}]U^{n+1} + d_{b}V^{n} \right]+\left[ [-d_{0}^{T}]U^{n} + d_{b}V^{n} \right]\right)
\\
+ \frac{1}{Bo}S^{n+1}
+\frac{2[\mu]_{\Gamma}}{Re}(N^{n+1*}+T^{n+1*}) + W_{\rho}^{n+1} G_{u}-\frac{1}{2}W_{\rho}^{n+1}d_{1}^{T}K^{n},
\end{split}\\
*_{2}d_{1}*_{1}^{-1} U^{n+1} = 0.
\end{gather}
where $N^{n+1*}$ and $T^{n+1*}$ are the vectors containing discrete normal strain rate dual 1-form and shear strain rate dual 1-form at $n+1*$ time step, which are defined as:
\begin{gather}
\left<\left(D_{nn}\nabla\phi\right)^{\flat},\star\sigma^{1}\right>^{n+1*}:=\sum_{\star\sigma^{2}\prec\star\sigma^{1}}\left(D_{nn}^{n}\nabla\phi^{n+1}\right)(\star\sigma^{2})\cdot(\star\vec{\sigma}^{1}),\\
\left<\left(D_{\tilde{t_{1}}n}\mathbf{t}_{1}\right)^{\flat},\star\sigma^{1}\right>^{n+1*}:=\sum_{\star\sigma^{2}\prec\star\sigma^{1}}\left(D_{\tilde{t_{1}}n}^{n}\mathbf{t}_{1}^{n+1}\right)(\star\sigma^{2})\cdot(\star\vec{\sigma}^{1}).
\end{gather}

Note that in the conservative PF method, $\phi = 0$ and $\phi =1$ distinguish the different pure phases. Therefore, the total mass of each phase can be computed as
\begin{equation}
m_{1}(t) = \rho_{1}m_{\phi}(t) = \sum\limits_{\sigma^{2}}\rho_{1}\phi(\star\sigma^{2},t)|\sigma^{2}|, \qquad\qquad m_{2}(t) = \rho_{2}m_{1-\phi}(t) = \sum\limits_{\sigma^{2}}\rho_{2}(1-\phi(\star\sigma^{2},t))|\sigma^{2}|,
\end{equation}
and the relative error of mass conservation of two phases is given by:
\begin{equation}
\varepsilon_{\phi}=\varepsilon_{1} = \frac{m_{1}(t)-m_{1}(0)}{m_{1}(0)}, \qquad\qquad \varepsilon_{1-\phi}=\varepsilon_{2}= \frac{m_{2}(t)-m_{2}(0)}{m_{2}(0)}.
\end{equation}

\section{Numerical results: advection test cases \label{verification}}
In this section, we verify the advection of the PF function for a prescribed velocity field. Test cases include the reversed single vortex test and Zalesak's disk test to evaluate the performance, such as boundedness, conservation and accuracy of the present DEC scheme. Also, a test case comprising advection of a bubble on a cylindrical surface is presented, which demonstrates that the present method is adaptable, without any changes, to curved domains.  The simulations are performed employing various regular/irregular meshes as summarized in Table \ref{Verification_tests_mesh_information_table}. The deviation of phase field state, from the initial state, at time $t$ can be defined as
\begin{eqnarray}
\delta_\Phi(t)=\sum\limits_{\sigma^{2}}|\phi(\star\sigma^{2},t)-\phi(\star\sigma^{2},t=0)||\sigma^{2}|.
\label{Difference_simulation}
\end{eqnarray}
As the following test cases are all time periodic with the exact phase field solution at $t=T_{i}$ to be the same as the initial condition, i.e., $\phi(0)=\phi(T_{i})$, the simulation error $\varepsilon$ is naturally defined as $\varepsilon=\delta_\Phi(T_{min})$, where $T_{min} = min\{T_{i}|\phi(T_{i})=\phi(0)\}$.

\begin{table}[htbp]
\caption{\label{tab1} Summary of the meshes employed for the advection verification tests.\protect \\
The mesh is identified by the convention \textit{$<Test\,name><Regular/Irregular><Number>$.
RSV, ZD and TC denote Reversed single vortex, Zalesak's disk and Translation
on cylinder, respectively. Regular meshes are indicated with ``R''
and irregular meshes with ``I''. $\left|\sigma^{1}\right|_{min},\left|\sigma^{1}\right|_{max}$
and $\left|\sigma^{1}\right|_{avg}$ are minimum, maximum and average
volume of 1-simplex, respectively. The number of 0-simplex,1-simplex
and 2-simplex are denoted by $N_{0},N_{1}$ and $N_{2}$, respectively.}}
\centering
\begin{tabular}{|lllllrrr|}
\hline

Test type & Mesh \newline\, name & $|\sigma^{1}|_{min}$ & $|\sigma^{1}|_{max}$ & $|\sigma^{1}|_{avg}$ & $N_{0}$ & $N_{1}$ & $N_{2}$ \\

\hline
\multicolumn{1}{|l|}{\multirow{6}{*}{Reversed single vortex}}& RSV-R1& 0.010101& 0.014285& 0.011486& 10000& 29601& 19602 \\
\cline{2-8}
\multicolumn{1}{|l|}{}&RSV-R2& 0.005030& 0.00711& 0.00572& 40000& 119201& 79202 \\
\cline{2-8}
\multicolumn{1}{|l|}{}&RSV-R3& 0.002506& 0.003544& 0.00285& 160000& 478401& 318402 \\
\cline{2-8}
\multicolumn{1}{|l|}{}&RSV-I1& 0.009220& 0.01739& 0.01296& 7039& 20806& 13768 \\
\cline{2-8}
\multicolumn{1}{|l|}{}&RSV-I2& 0.004072& 0.0074976& 0.005774& 34994& 104287& 69294 \\
\cline{2-8}
\multicolumn{1}{|l|}{}&RSV-I3& 0.001999& 0.0038706& 0.002992& 129685& 387716& 258032 \\\hline
\multicolumn{1}{|l|}{\multirow{6}{*}{Zalesak disk}} & ZD-R1& 0.010101& 0.014285& 0.011486& 10000& 29601& 19602  \\
\cline{2-8}
\multicolumn{1}{|l|}{}&ZD-R2& 0.005030& 0.00711& 0.00572& 40000& 119201& 79202 \\
\cline{2-8}
\multicolumn{1}{|l|}{}&ZD-R3& 0.002506& 0.003544& 0.00285& 160000& 478401& 318402 \\
\cline{2-8}
\multicolumn{1}{|l|}{}&ZD-I1& 0.007170& 0.01273& 0.00997& 11828& 35081& 26338 \\
\cline{2-8}
\multicolumn{1}{|l|}{}&ZD-I2& 0.004202& 0.00762& 0.00598& 32622& 97195& 64574 \\
\cline{2-8}
\multicolumn{1}{|l|}{}&ZD-I3& 0.002150& 0.00403& 0.00299& 129668& 387665& 257998 \\
\hline
\multicolumn{1}{|l|}{\multirow{4}{*}{Translation on cylinder}} &TC-R1& 0.019600& 0.0280& 0.0225& 32320& 96320& 64000 \\
\cline{2-8}
\multicolumn{1}{|l|}{}&TC-R2& 0.009820& 0.01401& 0.0113& 128640& 384640& 256000 \\
\cline{2-8}
\multicolumn{1}{|l|}{}&TC-I1& 0.019000& 0.0357& 0.0266& 20732& 61724& 40992 \\
\cline{2-8}
\multicolumn{1}{|l|}{}&TC-I2& 0.009640& 0.0173& 0.0134& 81140& 242484& 161344 \\
\hline
\end{tabular}
\label{Verification_tests_mesh_information_table}
\end{table}

\subsection{Reversed single vortex test}
\begin{figure}[htbp]
\centering
\includegraphics[width=0.8\linewidth]{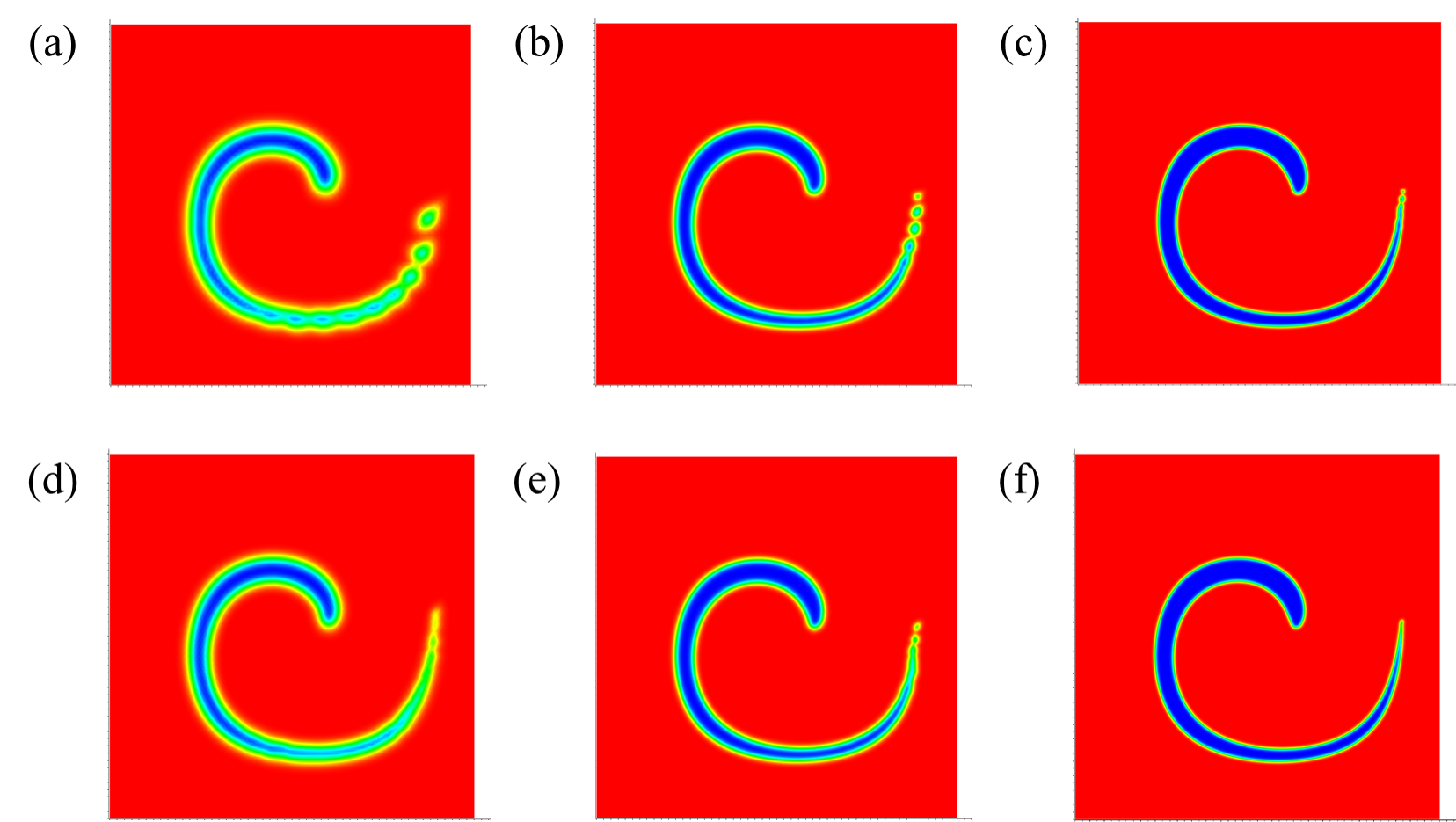}
\caption{The interface profiles of reversed single vortex tests at $t=T/2$ on different meshes. Regular mesh solutions are shown in 
(a) RSV-R1, (b) RSV-R2, (c) RSV-R3, while irregular mesh solutions are in (d) RSV-I1, (e) RSV-I2, (f) RSV-I3.
See Table \ref{Verification_tests_mesh_information_table} for the detailed mesh information.
 }
\label{Reversed_vortex_half_T}
\end{figure}

\begin{figure}[htbp]
\centering
\includegraphics[width=0.8\linewidth]{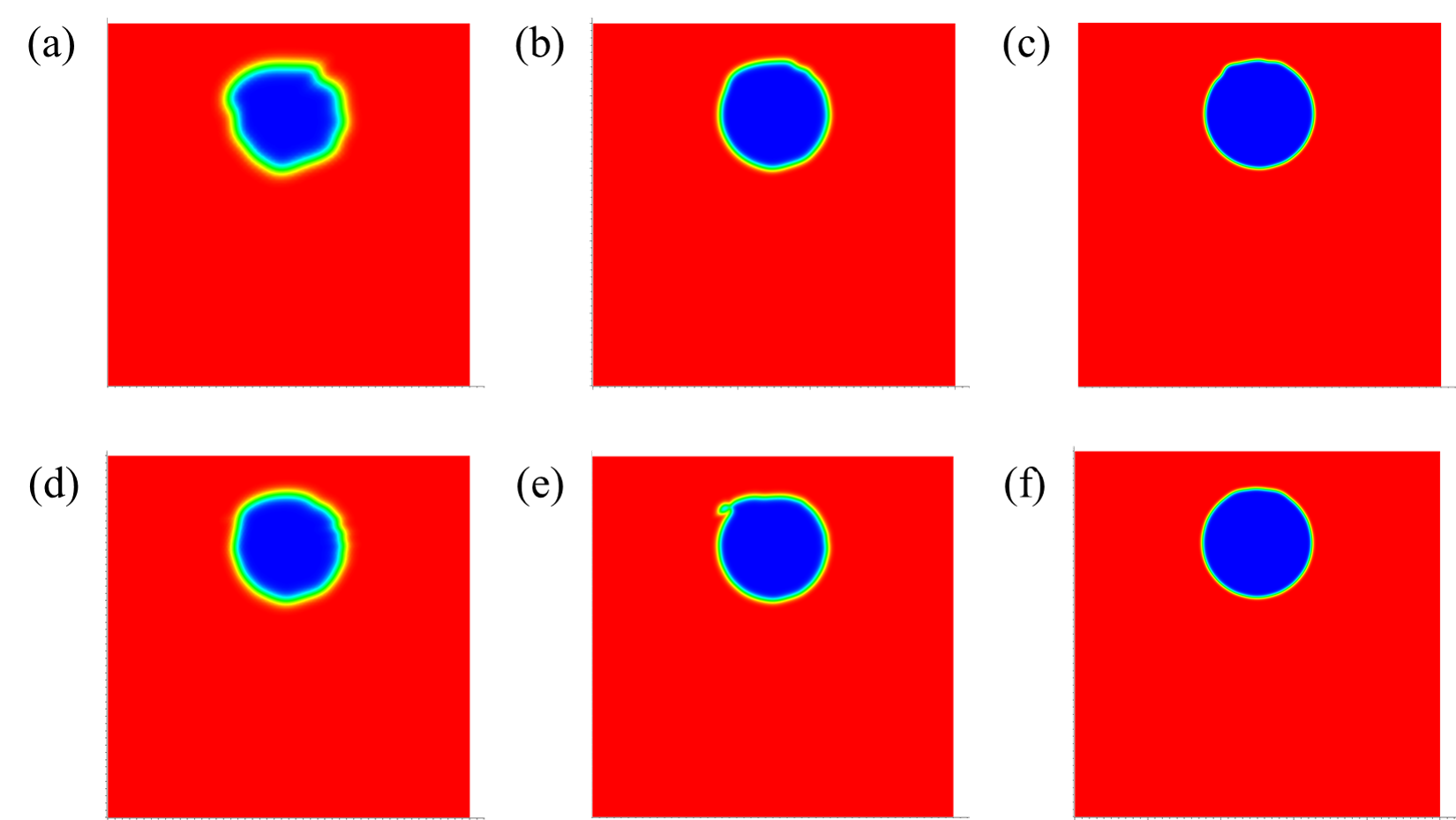}
\caption{The interface profiles of reversed single vortex tests at $t=T$ on different meshes. Regular mesh solutions are shown in 
(a) RSV-R1, (b) RSV-R2, and (c) RSV-R3, while irregular mesh solutions are shown in (d) RSV-I1, (e) RSV-I2, and (f) RSV-I3.
See Table \ref{Verification_tests_mesh_information_table} for the detailed mesh information. }
\label{Reversed_vortex_T}
\end{figure}

\begin{figure}[htbp]
\centering
\includegraphics[width=0.8\linewidth]{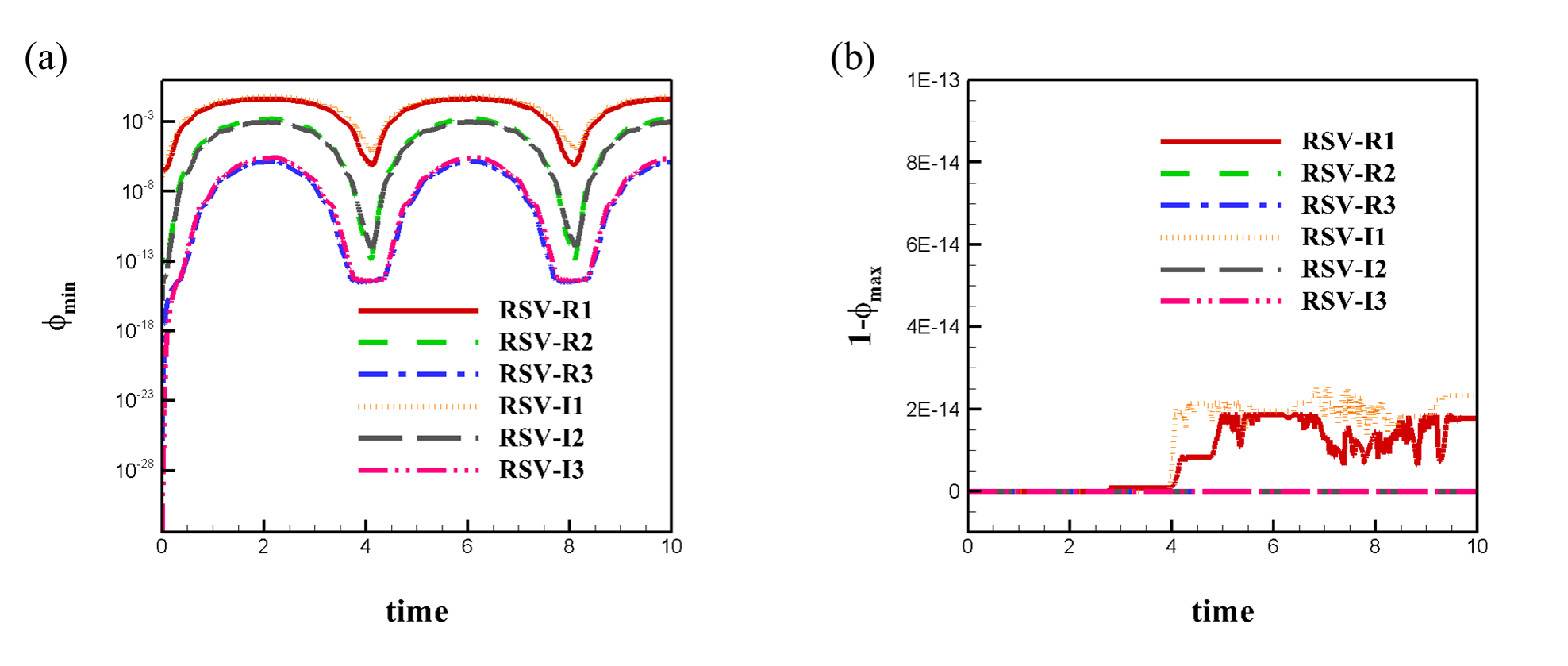}
\caption{The bounds of $\phi$ in reversed single vortex test: (a) lower bound (global minimum); (b) upper bound (global maximum). }
\label{Reversed_vortex_boundedness}
\end{figure}

\begin{figure}[htbp]
\centering
\includegraphics[width=0.8\linewidth]{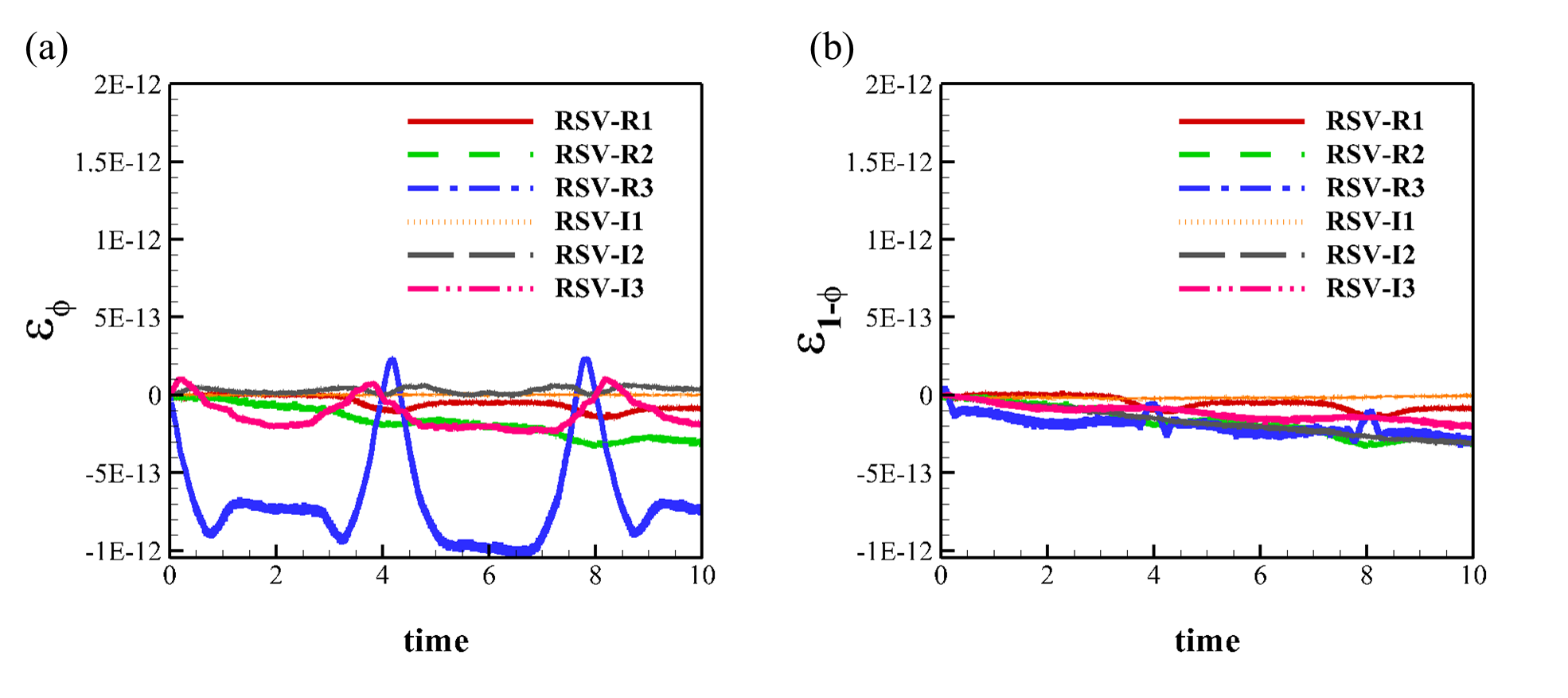}
\caption{The mass conservation error for reversed single vortex test: (a) one phase, (b) the other phase.}
\label{Reversed_vortex_mass_conservation}
\end{figure}

As a standard test, reversed single vortex test, originally introduced by Rider \etal \citep{rider1998reconstructing}, is suitable for verifying the accuracy of a numerical advection scheme. In this simulation, a circular patch with radius $R=0.15$ is centered at $(x,y)=(0.5,0.75)$ in a unit square  $[0,1]\times[0,1]$, initially. The circular patch is advected and deformed by the velocity field which is given by the following stream function
\begin{eqnarray}
\psi(x,y,t)=\frac{1}{\pi}\sin^{2}(\pi x)\sin^{2}(\pi y)\cos(\frac{\pi t}{T}),
\label{Stream_function_reversed}
\end{eqnarray}
where the stream function period is $2T=8$. The time periodicity of the stream function implies that the velocity field also  varies periodically in time, which stretches and contracts the circular patch in a manner that after every time interval of duration equal to half the period, $T$,  the patch ideally returns to its original configuration. In the first quarter of the time period, the velocity field elongates the circle, followed by a  contraction it in the second quarter of the time period until the circle returns to the initial state at $t=T$. The motion state in the second half time period is a reflection about the straight line $x = 0.5$, which is parallel to $y-$ axis, of the solution in the first half period. 
Two different sets meshes were employed: regular and irregular. For each set, three different mesh resolutions were used. See Table \ref{Verification_tests_mesh_information_table} for the detailed mesh information with names prefixed with ``RSV'. Figures \ref{Reversed_vortex_half_T} and \ref{Reversed_vortex_T}
show the shape of fully stretched state $(t = T/2)$ and the end reversed state $(t=T)$, respectively. Figure \ref{Reversed_vortex_boundedness} shows the lower bound and upper bounds of $\phi$. For a choice of free parameters $\gamma$ and $\epsilon$ as discussed in section \ref{section_boundedness_analysis}, the present scheme can keep the boundedness of PF function $\phi$ without any special treatment such as redistribution of mass. Figure \ref{Reversed_vortex_mass_conservation} shows the mass error of the two phases as a function of time. The total mass of the two phases is conserved to the machine precision.  As mentioned before, the exact solution at $t=T$ is the same as the initial state. The  simulation error (see equation (\ref{Difference_simulation})) is reported in Table \ref{Reversed_vortex_error_table} for different meshes. As expected, the numerical results converge towards exact solution as the mesh resolution increases.

\begin{table}[htbp]
\caption{\label{tab2}Error of the reversed single vortex test for different meshes. RSV-R (resp. RSV-I) indicates the use of a regular (resp. irregular) mesh.}
\centering
\begin{tabular}{|ccccccc|}
\hline

Mesh \newline\, name & Regular RSV-R1 & RSV-R2 &RSV-R3 & RSV-I1 & RSV-I2 &  RSV-I3 \\

\hline
Error $\varepsilon$& 6.61e-03& 1.42e-03& 6.96e-04& 8.77e-03& 2.86e-03& 7.72e-04\\
\hline
\end{tabular}
\label{Reversed_vortex_error_table}
\end{table}


\subsection{Zalesak's disk test}
\begin{figure}[htbp]
\centering
\includegraphics[width=0.8\linewidth]{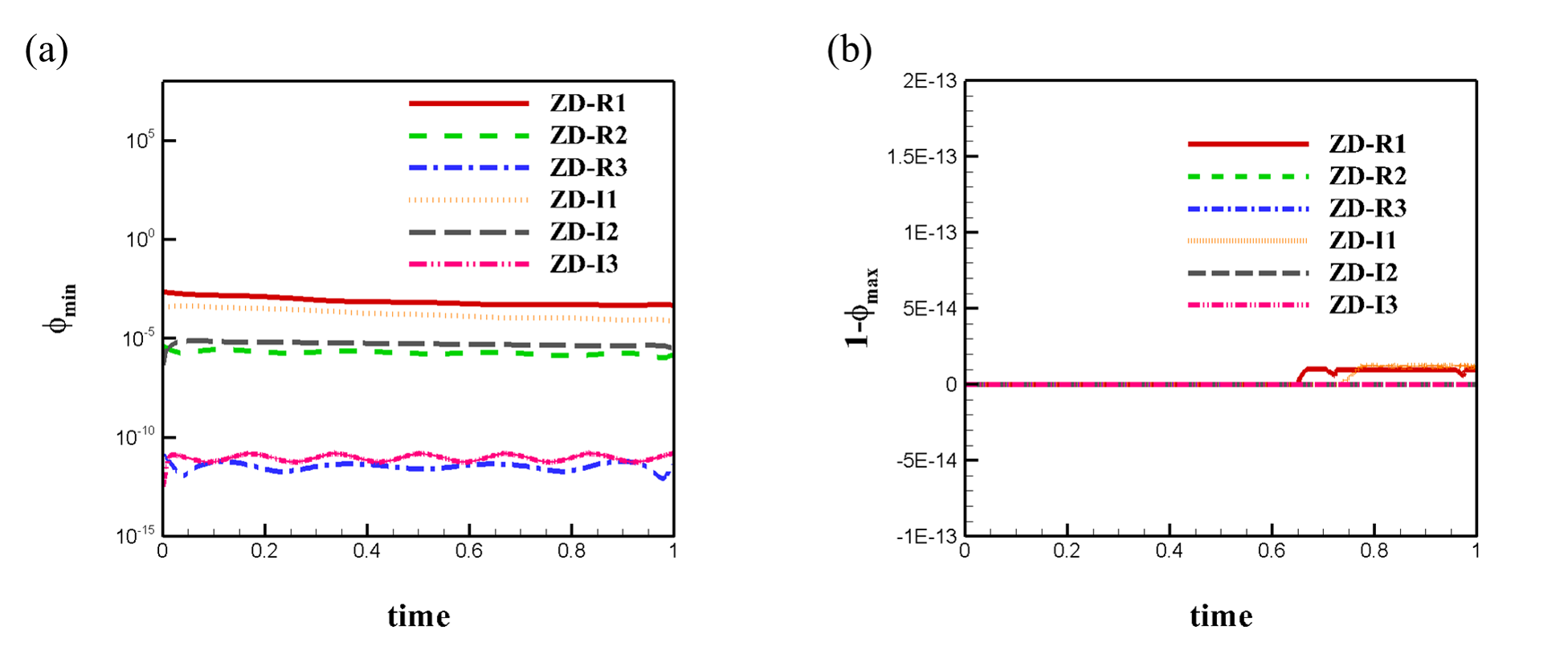}
\caption{The bounds of $\phi$ in the Zalesak disk test: (a) lower bound (global minimum); (b) upper bound (global maximum).}
\label{Zalesak_disk_boundedness}
\end{figure}

\begin{figure}[htbp]
\centering
\includegraphics[width=0.8\linewidth]{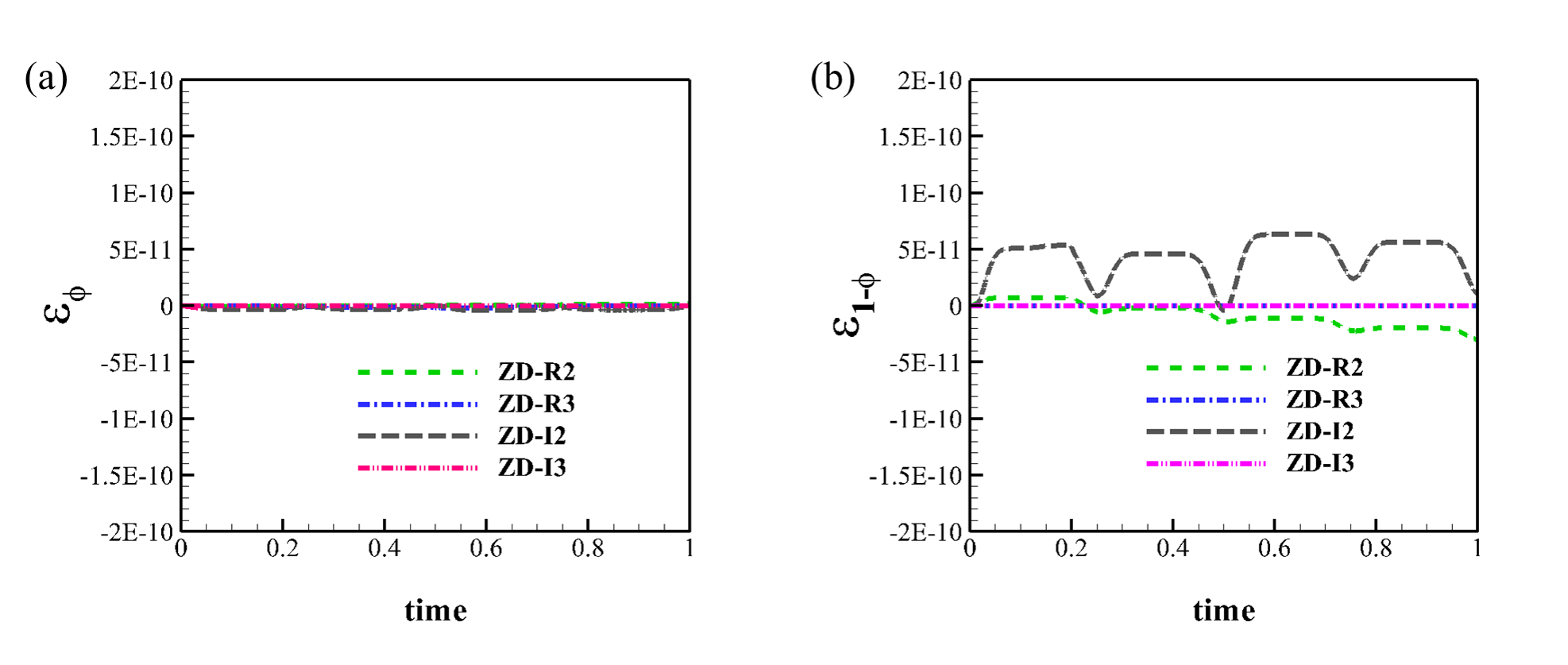}
\caption{Mass conservation error for the Zalesak disk test: (a) one phase, (b) the other phase.}
\label{Zalesak_disk_mass_conservation}
\end{figure}

\begin{figure}[htbp]
\centering
\includegraphics[width=0.8\linewidth]{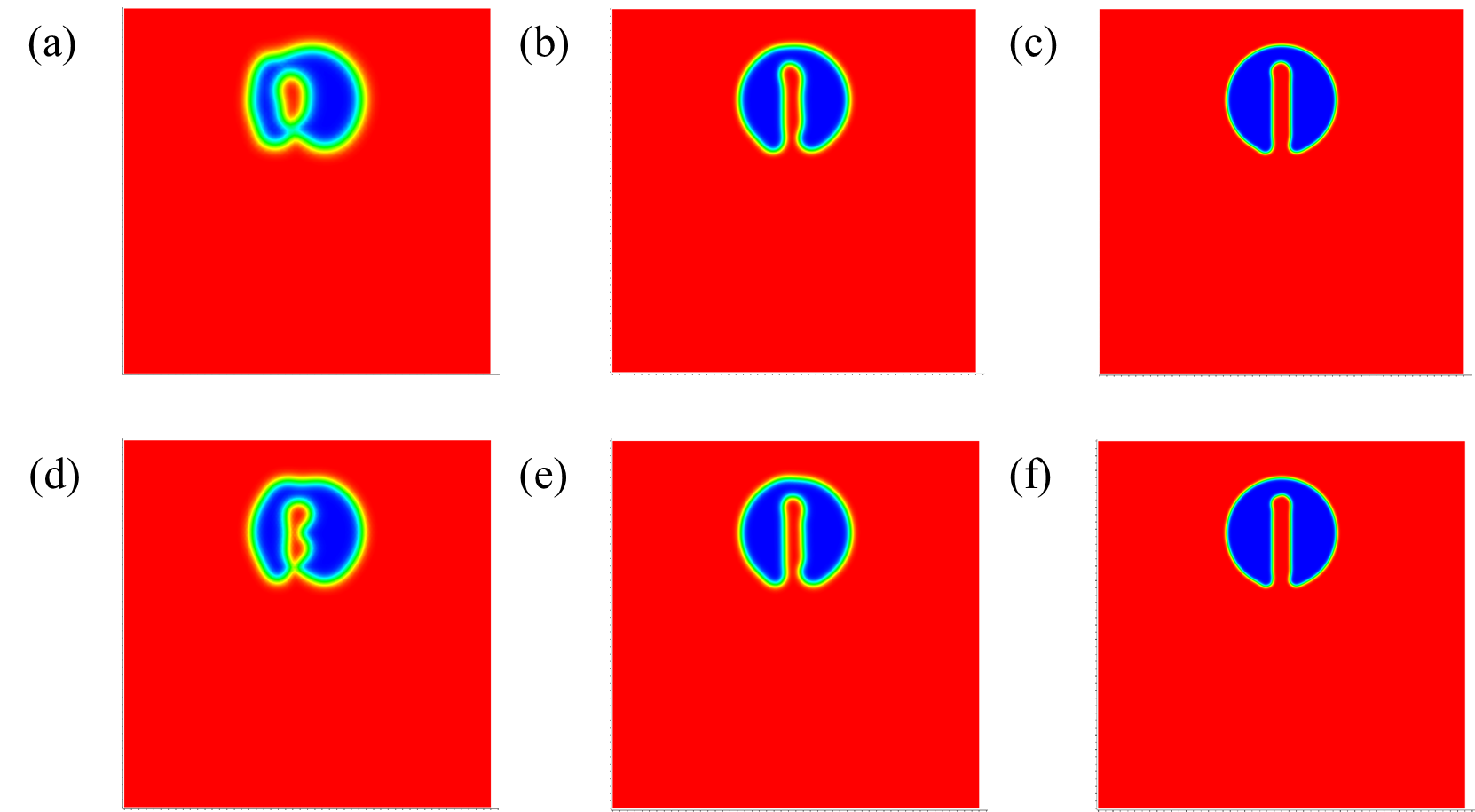}
\caption{The interface profiles of the Zalesak disk tests at $t=T$ for different meshes. Regular mesh solutions are shown in  (a)ZD-R1, (b) ZD-R2, and (c) Regular-ZD-R3; whereas irregular mesh solutions are shown in  (d) ZD-I1, (e) ZD-I2, and (f) ZD-I3. See Table \ref{Verification_tests_mesh_information_table} for the detailed mesh information.}
\label{Zalesak_disk_T}
\end{figure}

The test case comprising solid body rotation of a slotted disk, also called Zalesak's disk test \citep{zalesak1979fully}, is conventionally used for evaluating the performance of advection schemes. In this test, a slotted disk of radius $R=0.15$ with slot width $h=0.05$ and length $l=0.25 $ is centered at $(0, 0.25)$ of a unit square domain $[-0.5,0.5]\times[-0.5,0.5]$ initially. The solid body rotation of the Zalesak's disk is subjected to the following velocity field
\begin{eqnarray}
u = -2\pi y, \qquad v = 2\pi x.
\label{velocity_field_slotted}
\end{eqnarray}
Analytically, the disk maintains its shape and does time periodic motion, due to the application of a velocity field resembling solid body rotation. It recovers the initial condition after one time period $T=1$. Two different sets meshes were employed: regular and irregular. For each set, three different mesh resolutions were used. See Table \ref{Verification_tests_mesh_information_table} for the detailed mesh information with names prefixed with ``ZD''.  The boundedness of PF function and total mass conservation for the two phases are shown in Figure \ref{Zalesak_disk_boundedness} and Figure \ref{Zalesak_disk_mass_conservation}, respectively. Mass conserves to the machine precision for the finer meshes. There is some loss of mass conservation for coarse meshes owing to the fact that the domain boundaries are not impermeable and fluid can flux out of the domain.  Figure \ref{Zalesak_disk_T} shows the shape of slotted disk at the end state (at the end of a time period) for different meshes, which reveals  that the simulation results are closer to exact solution for the finer mesh. The simulation error $\varepsilon=\delta_\Phi(T)$, as defined in equation (\ref{Difference_simulation}), is reported in Table \ref{Zalesak_disk_error_table} for different meshes.

\begin{table}[htbp]
\caption{\label{tab3}Error of the Zalesak disk tests for different meshes}
\centering
\begin{tabular}{|ccccccc|}
\hline

Mesh \newline\, name & ZD-R1 &ZD-R2 & ZD-R3  & ZD-I1 & ZD-I2 & ZD-I3 \\

\hline
Error $\varepsilon$& 1.86e-02& 3.66e-03& 7.91e-04& 1.45e-02& 4.60e-03& 1.03e-03\\
\hline
\end{tabular}
\label{Zalesak_disk_error_table}
\end{table}

\subsection{Circle advection on a curved (cylindrical) surface}
\begin{figure}[htbp]
\centering
\includegraphics[width=0.95\linewidth]{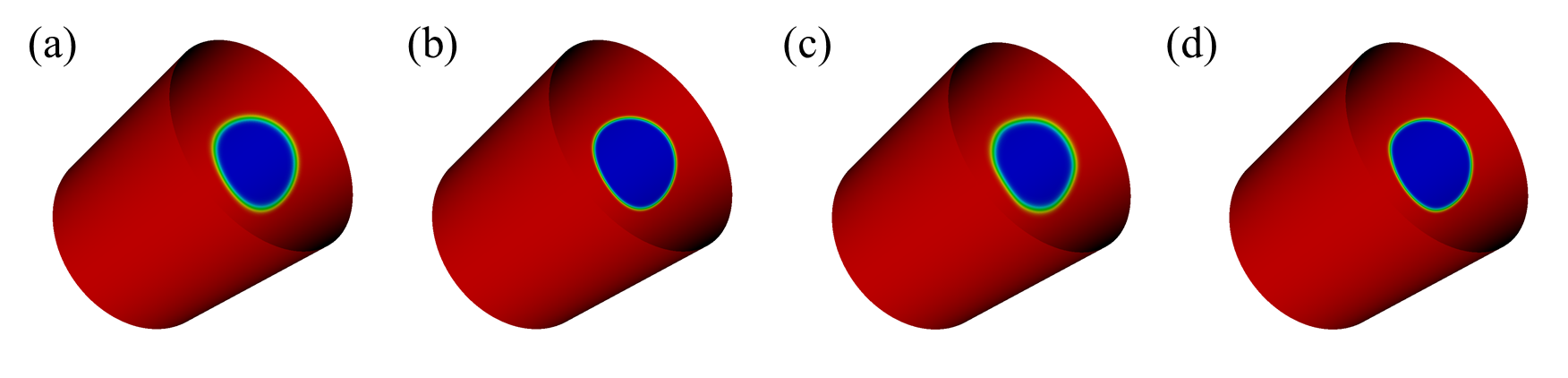}
\caption{The interface profiles for the circle translation on a cylinder at $t=T/2$ for different meshes. Regular mesh solutions are shown in  (a) TC-R1, and (b) TC-R2; whereas irregular mesh solutions are shown in (c) Irregular-TC-I1, and (d) Irregular-TC-I2. See Table \ref{Verification_tests_mesh_information_table} for the detailed mesh information.}
\label{Cylinder_convection_half_T}
\end{figure}

\begin{figure}[htbp]
\centering
\includegraphics[width=0.95\linewidth]{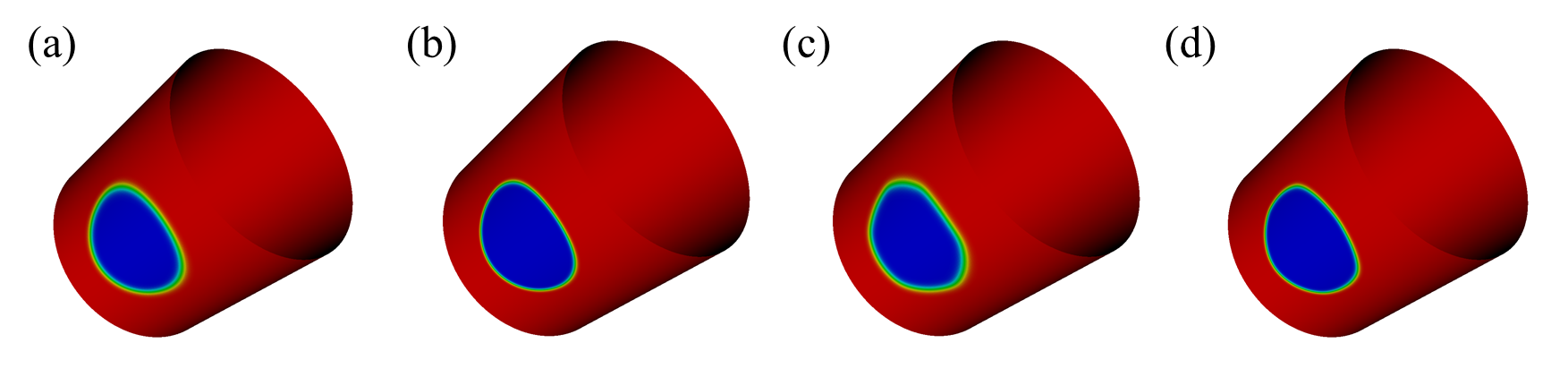}
\caption{The interface profiles for the circle translation on a cylinder at $t=T$ for different meshes. Regular mesh solutions are shown in (a) TC-R1, and (b) TC-R2; whereas irregular mesh solutions are shown in (c) TC-I1, and (d) TC-I2.}
\label{Cylinder_convection_T}
\end{figure}

\begin{figure}[htbp]
\centering
\includegraphics[width=0.8\linewidth]{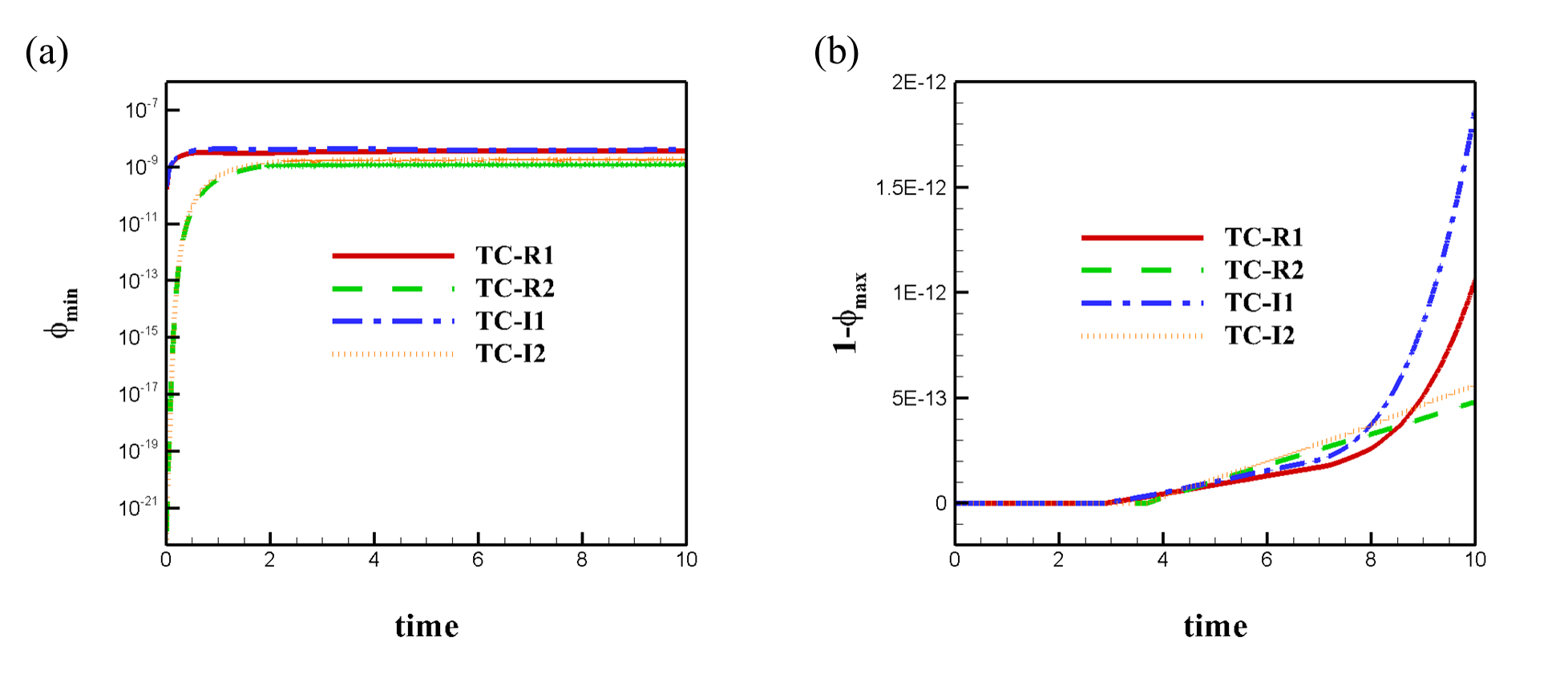}
\caption{Bounds of $\phi$ for the circle advection test on a cylindrical surface: (a) lower bound (global minimum); (b) upper bound global maximum.}
\label{Cylinder_convection_boundedness}
\end{figure}

\begin{figure}[htbp]
\centering
\includegraphics[width=0.8\linewidth]{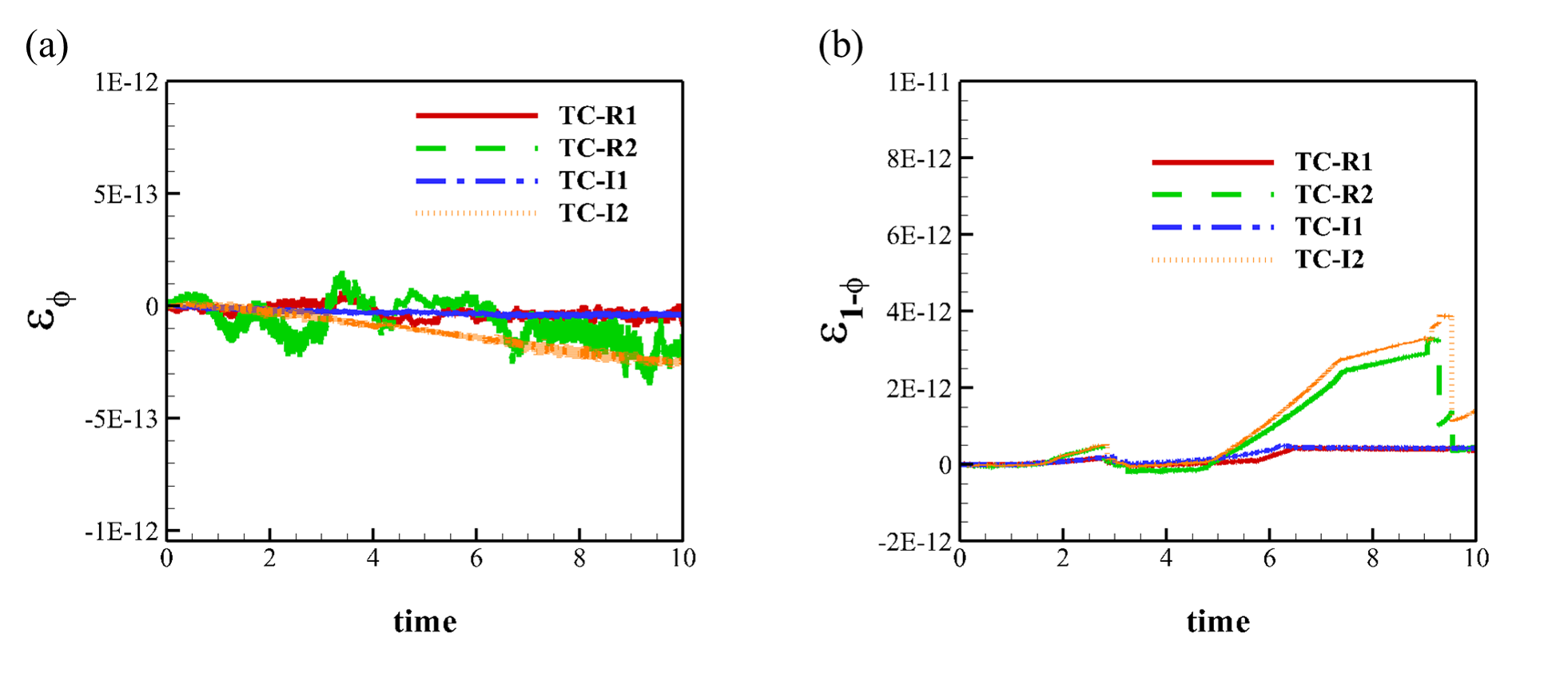}
\caption{ Mass conservation error for the circle advection test on a cylindrical surface: (a) one phase, (b) the other phase.}
\label{Cylinder_convection_mass_conservation}
\end{figure}

The uniform translation of a circular patch under the influence of a constant velocity field is considered an important verification test  for a discrete advection scheme. Usually, the domain is a flat surface for this test. However, to demonstrate the adaptability of our method to the curved domains, we present here the analogous test on a curved surface. A circle with radius $R=0.5$ is embedded on a compact 2D Riemannian manifold comprising a full cylinder (height $h=2$, radius $r=1$) surface with unit velocity around it in this simulation.  See Table \ref{Verification_tests_mesh_information_table} for the detailed mesh information with names prefixed with ``TC''. The circle advects with a uniform surface speed with time period $T=2\pi$. The solutions at $t=T/2$ and $t=T$ are shown in Figures \ref{Cylinder_convection_half_T} and \ref{Cylinder_convection_T}, respectively, for different meshes. Figures \ref{Cylinder_convection_boundedness} and  \ref{Cylinder_convection_mass_conservation} show the boundedness of PF function and exact mass conservation, respectively.  The simulation error $\varepsilon=\delta_\Phi(T)$  for different meshes is reported in Table \ref{Cylinder_convection_error_table}.

\begin{table}[htbp]
\caption{\label{tab4}Error of circle translation on a cylinder for different meshes}
\centering
\begin{tabular}{|ccccc|}
\hline

Mesh \newline\, name & TC-R1 & TC-R2 & TC-I1 & TC-I2 \\

\hline
Error $\varepsilon$ & 4.89e-03& 2.32e-03& 1.76e-02& 8.80e-03\\
\hline
\end{tabular}
\label{Cylinder_convection_error_table}
\end{table}

\section{Numerical results:  two-phase Navier-Stokes simulations \label{physical_phenomena}}
In this section, we present the simulation results with the full two-phase Navier-Stokes equation (comprising both, the incompressible Navier-Stokes and conservative phase field equations) for various physical phenomena such as the Rayleigh-Taylor instability, a dam break problem, a rising bubble, and a free falling drop on a flat pool surface.  Moreover, we augment the above test cases with that of a rising bubble on full and half cylinders to investigate and explore the bubble dynamical evolution on a curved surface. In these simulations, the free slip and Neumann boundary conditions were applied for velocity and phase field, respectively. Table \ref{Numerical_results_mesh_information_table} shows the summary of the meshes employed for the simulations.
%
%
\begin{table}[htbp]
\caption{\label{tab5}Mesh information for two-phase Navier-Stokes simulation test cases presented in section \ref{physical_phenomena}. All meshes are of the regular type.}
\centering
\begin{tabular}{|lllllrrr|}
\hline

Test type & Mesh \newline\, name & $|\sigma^{1}|_{min}$ & $|\sigma^{1}|_{max}$ & $|\sigma^{1}|_{avg}$ & $N_{0}$ & $N_{1}$ & $N_{2}$ \\

\hline
\multicolumn{1}{|l|}{\multirow{1}{*}{Rayleigh-Taylor instability}} & RTI& 0.01& 0.01414& 0.01137& 40501& 120500& 80000 \\
\hline
 \multicolumn{1}{|l|}{\multirow{1}{*}{Dam break}}& DB& 0.010101& 0.01428& 0.0115& 10000& 29601& 19602 \\
\hline
\multicolumn{1}{|l|}{\multirow{3}{*}{Rising bubble}}& RB1& 0.08& 0.11552& 0.09271& 1275& 3674& 2400 \\
\cline{2-8}
\multicolumn{1}{|l|}{}&RB2& 0.04& 0.05657& 0.04547& 5151& 15150& 10000\\
\cline{2-8}
\multicolumn{1}{|l|}{}&RB3& 0.02& 0.028284& 0.02275& 20301& 60300& 40000\\
\hline
\multicolumn{1}{|l|}{\multirow{1}{*}{Drop impact}} & DP& 0.08& 0.1131& 0.0909& 5151& 15150& 10000\\
\hline
\multicolumn{1}{|l|}{\multirow{1}{*}{Bubble on a half cylinder}}& HC& 0.07853& 0.1121& 0.0900& 4131& 12130& 8000\\
\hline
\multicolumn{1}{|l|}{\multirow{1}{*}{Bubble on a flat plane}}&FP& 0.04& 0.05657& 0.04547& 5151& 15150& 10000 \\
\hline
\multicolumn{1}{|l|}{\multirow{1}{*}{Bubble on a full cylinder}}&FC& 0.07853& 0.1121& 0.0901& 8160& 24160& 16000 \\
\hline
\end{tabular}
\label{Numerical_results_mesh_information_table}
\end{table}

\subsection{Rayleigh-Taylor instability}
\begin{figure}[htbp]
\centering
\includegraphics[width=0.8\linewidth]{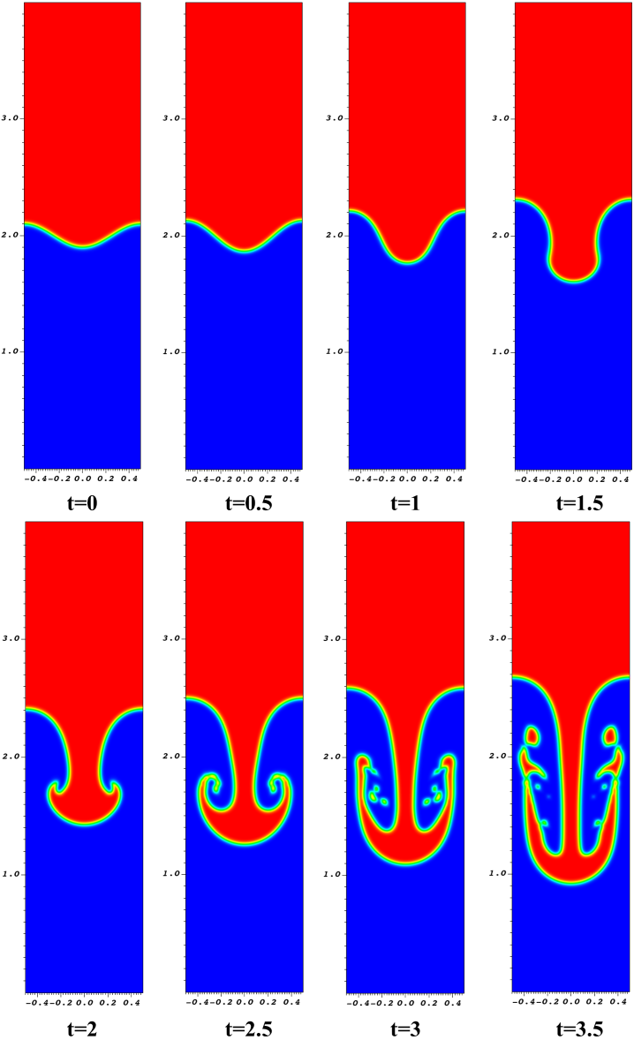}
\caption{The interface evolution of Rayleigh-Taylor instability.}
\label{RTI_t_evolution}
\end{figure}

\begin{figure}[htbp]
\centering
\includegraphics[width=0.5\linewidth]{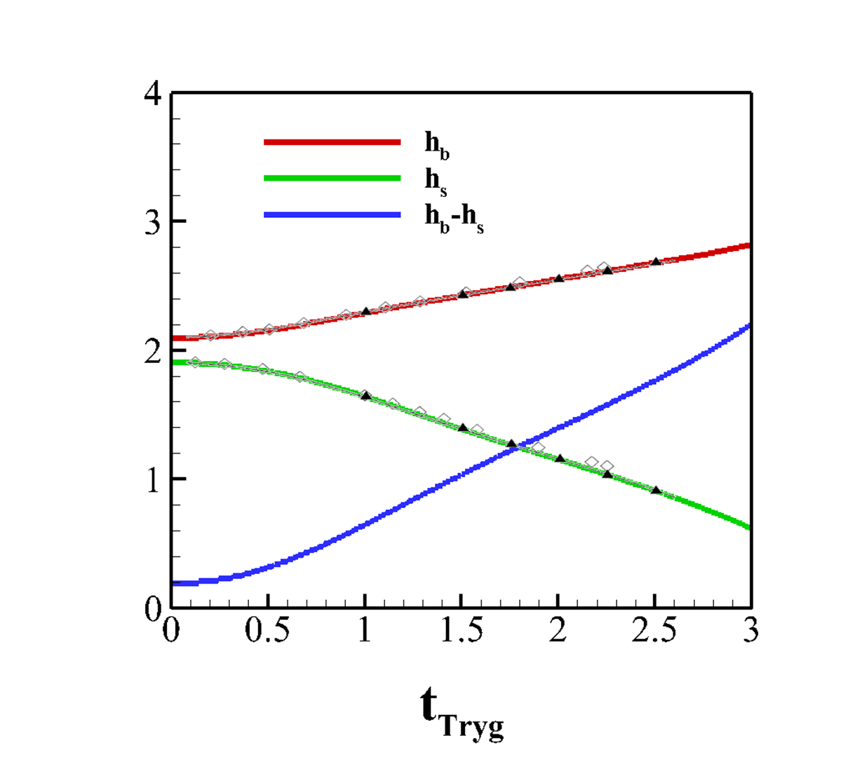}
\caption{The evolution of bubble height $h_{b}$, spike height $h_{s}$ and the height difference $h_{b} - h_{s}$. The black filled triangles and open diamonds \citep{ding2007diffuse} correspond to the solution of Guermond \citep{guermond2000projection} and Tryggvason \citep{tryggvason1988numerical}.}
\label{RTI_h_b_h_s_Tryg}
\end{figure}

The Rayleigh-Taylor instability (RTI) in the instability of an interface separating fluids of different densities under the influence of a gravitational field. It is characterized by the penetration of the heavier fluid into the lighter fluid (aka ``spikes'') and vice versa (aka ``bubbles"). 
A non-dimensional parameter which characterizes the density difference between the two phases is Atwood number, $At = (\rho_{H}-\rho_{L})/(\rho_{H}+\rho_{L})$, where $\rho_{H}$ and $\rho_{L}$ are the density of heavy and light fluids, respectively. We consider RTI configurations similar to those in  some previous works  \citep{guermond2000projection, ding2007diffuse, chiu2011conservative}, i.e.,the Atwood number $(At)$, Reynolds number $(Re)$ and viscosity ratio $\mu_{H}/\mu_{L}$ are chosen to be $0.5$, $3000$ and $1$, respectively. The initial perturbed interface is in a rectangular domain $[-D/2,D/2]\times[0,4D]$, and the interface is defined as $y(x) = 2D - 0.1D\cos(2\pi x/D)$. For this test case, the surface tension is neglected $(\sigma = 0)$. The length, density and viscosity scales are $D$, $\rho_{H}$, and $\mu_{H}$, respectively. Hence, the dimensionless parameters are $\rho_{r}= 1/3$, $\mu_{r}=1$, $Re=3000$ and $Bo=\infty$. Table \ref{Numerical_results_mesh_information_table} provides the mesh information. The interface evolution is given in Figure \ref{RTI_t_evolution}. The time histories of bubble height $h_{b}$ / spike height $h_{s}$ and their height difference $h_{b} - h_{s}$, where time is normalized by the Tryggvason time scale $(t_{Tryg}=t\sqrt{At})$ are shown in Figure \ref{RTI_h_b_h_s_Tryg}. Our results are in good agreement with those reported in references  \citep{tryggvason1988numerical,guermond2000projection,ding2007diffuse}

\subsection{Dam-break problem}
The dam break simulation is a classical and well researched problem in the field of two-phase flows \citep{sun2010coupled, chiu2011conservative, patel2017novel, xie2020conservative}. It generally involves large interface distortions. In our simulation, a liquid column of length $L_{l}=L$ and width $W_{l}=L/2$ is located in the left bottom of a tank of length $L_{t}=2L$ and width $W_{t}=2L$ initially. The reference length is chosen as $2L$.  The mesh information is provided in Table \ref{Numerical_results_mesh_information_table}. We perform two simulations:  (a) High Bond number case in which the fluid properties are the same as in some pervious works \citep{martin1952experimental, sun2010coupled,chiu2011conservative}, i.e., the parameters are chosen as density ratio $\rho_{r}$, viscosity ratio $\mu_{r}$, Reynolds number $Re$ and Bond number $Bo$ are $1000$, $1000$, $2794.23$ and $44.27$, respectively; and (b) Low Bond number case in which the dimensionless parameters are set as $\rho_{r}=1000$, $\mu_{r}=100$ $Re=10$ and $Bo=1/800$. 

\begin{figure}[htbp]
\centering
\includegraphics[width=0.9\linewidth]{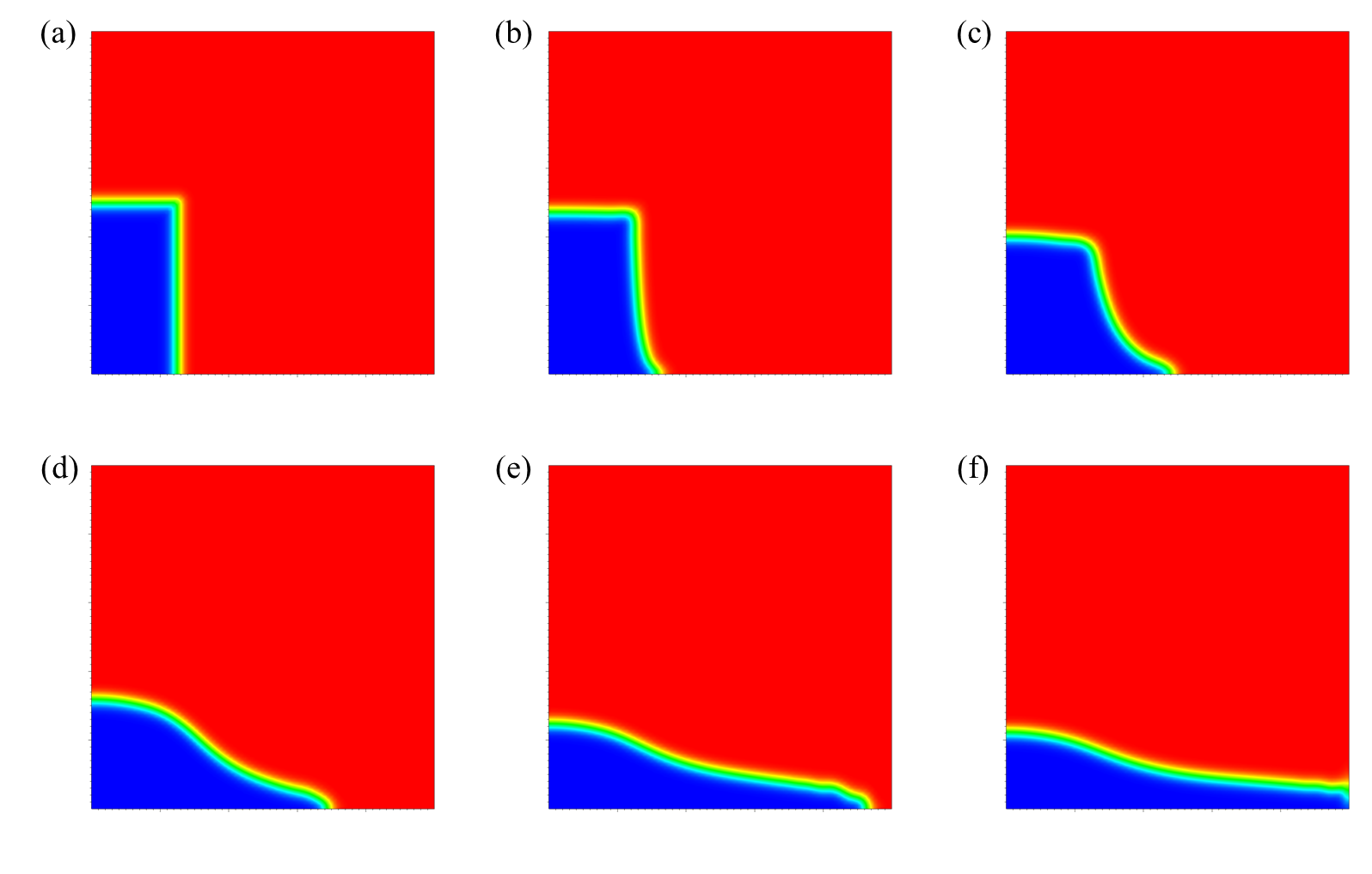}
\caption{The interface evolution of the low surface tension Dam-break simulation with $\rho_{r} = 1000$, $\mu_{r} = 1000$, $Re = 2794.23$ and $Bo = 44.27$ shown at various times: (a) $t=0$, (b) $t=0.25$, (c) $t=0.5$, (d) $t=0.75$, (e) $t=1$, (f) $t=1.1$.}
\label{Dam_break_2_evolution}
\end{figure}

\begin{figure}[htbp]
\centering
\includegraphics[width=0.5\linewidth]{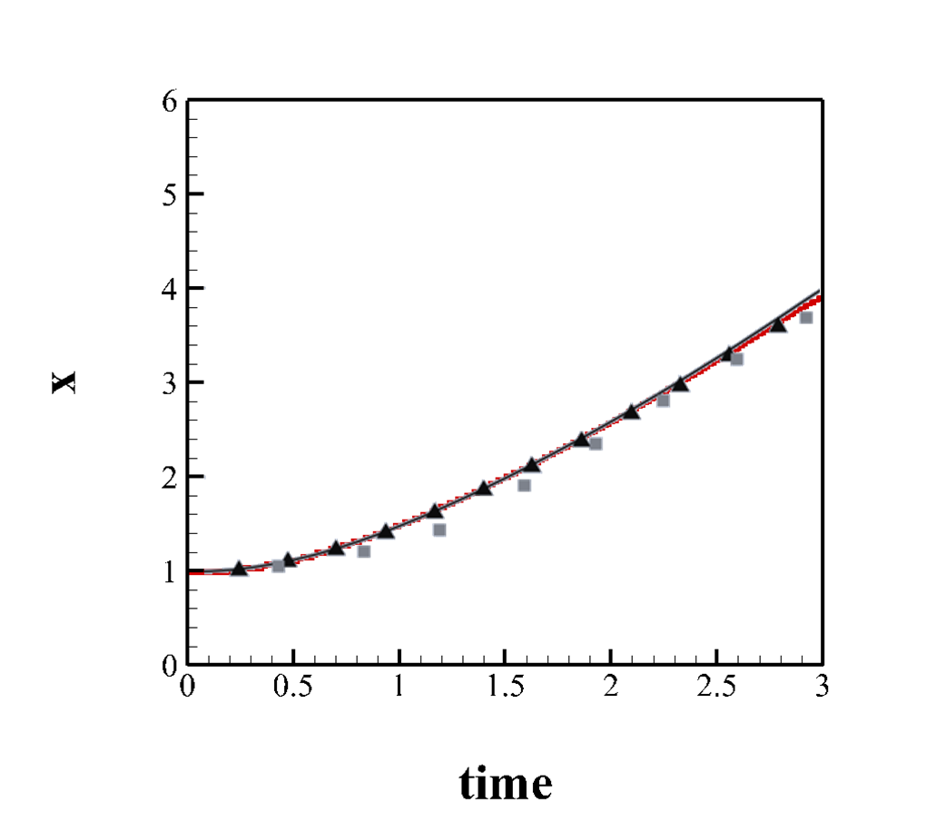}
\caption{The time history of the liquid right front for the low surface tension Dam-break problem. The black filled triangles, gray filled square, black solid line and red solid line correspond to the result of Sun \citep{sun2010coupled}, Martin \citep{martin1952experimental}, Chiu \citep{chiu2011conservative} and current simulation, respectively.
}
\label{Dam_break_right_edge}
\end{figure}

\begin{figure}[htbp]
\centering
\includegraphics[width=0.9\linewidth]{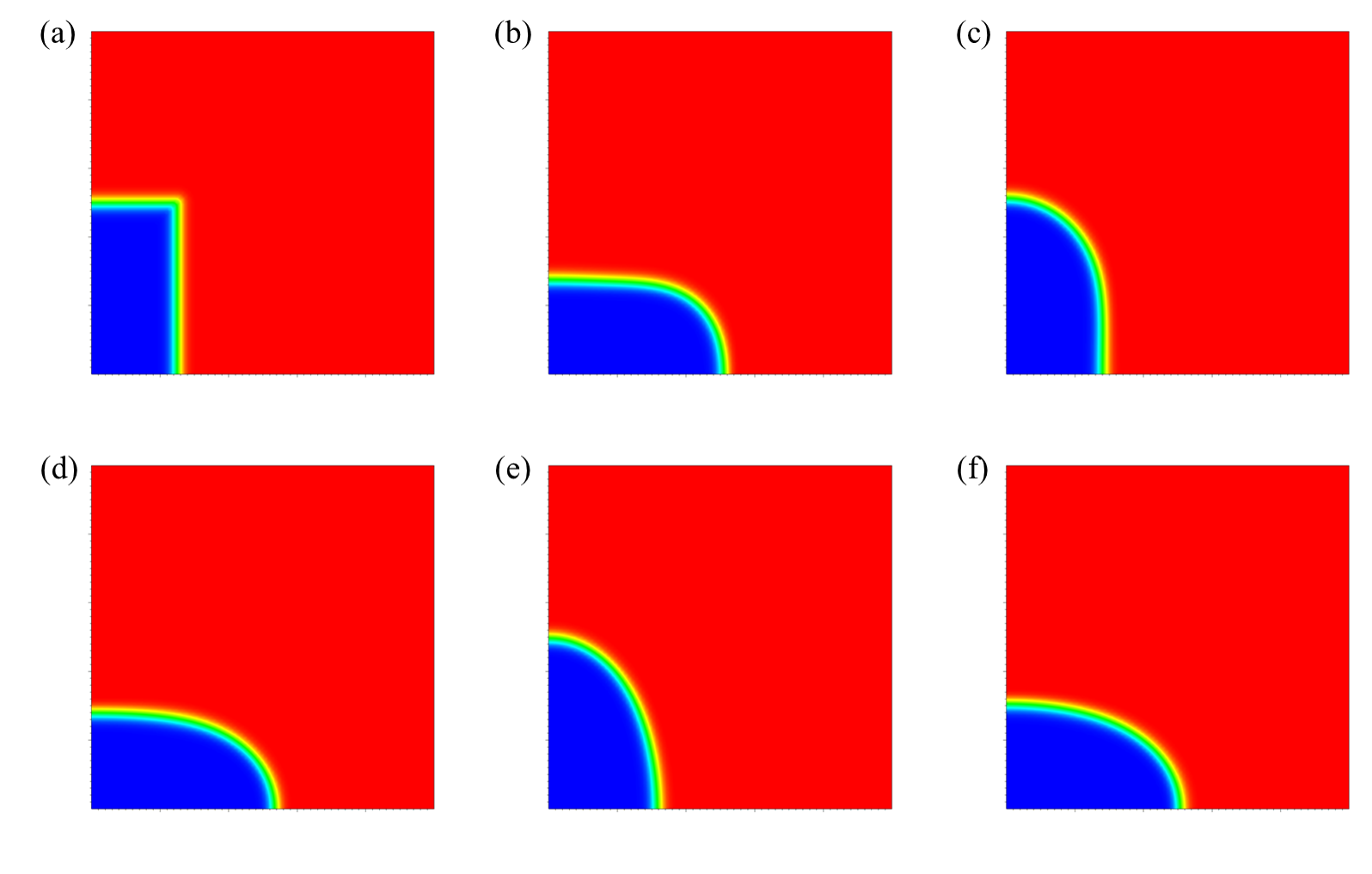}
\caption{The interface evolution of the Dam-break simulation with $\rho_{r} = 1000$, $\mu_{r} = 100$, $Re = 10$ and $Bo = 1/800$: (a) $t=0$, (b) $t=0.5$, (c) $t=0.8$, (d) $t=1.2$, (e) $t=1.6$, (f) $t=1.9$.}
\label{Dam_break_1_2_evolution}
\end{figure}

\begin{figure}[htbp]
\centering
\includegraphics[width=0.5\linewidth]{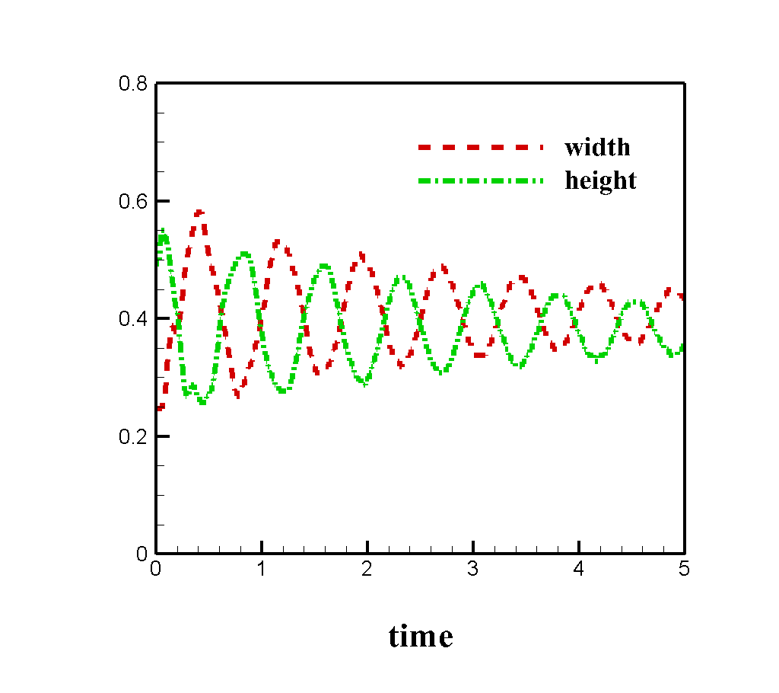}
\caption{The time history of the liquid height and width for the high surface tension Dam-break problem. 
}
\label{Dam_break_right_left_edge}
\end{figure}

For the high Bond number case, the interface evolution is shown in Figure \ref{Dam_break_2_evolution}. Recording the history of fluid front and transforming to the length and time scale given in \citep{sun2010coupled,chiu2011conservative}, Figure \ref{Dam_break_right_edge} shows our numerical result agrees very well with previous work \citep{martin1952experimental,sun2010coupled,chiu2011conservative}. 
For the low Bond number case,  the surface tension is larger as compared to the former simulation. 
Thus, in this simulation,  the large surface tension counteracts gravity and forces back the liquid, which leads to oscillations of the interface. Figure \ref{Dam_break_1_2_evolution} shows that the liquid oscillation under large surface tension and the oscillation amplitude decays due to viscosity. Figure \ref{Dam_break_right_left_edge} shows the height and width (the maximum distance from the interface to $x=0$ and $y=0$, respectively) of liquid as a function of time, from which we compute the frequency of oscillation of the liquid to be  $1.3333$.

\subsection{Rising bubble problem}
\begin{figure}[htbp]
\centering
\includegraphics[width=0.9\linewidth]{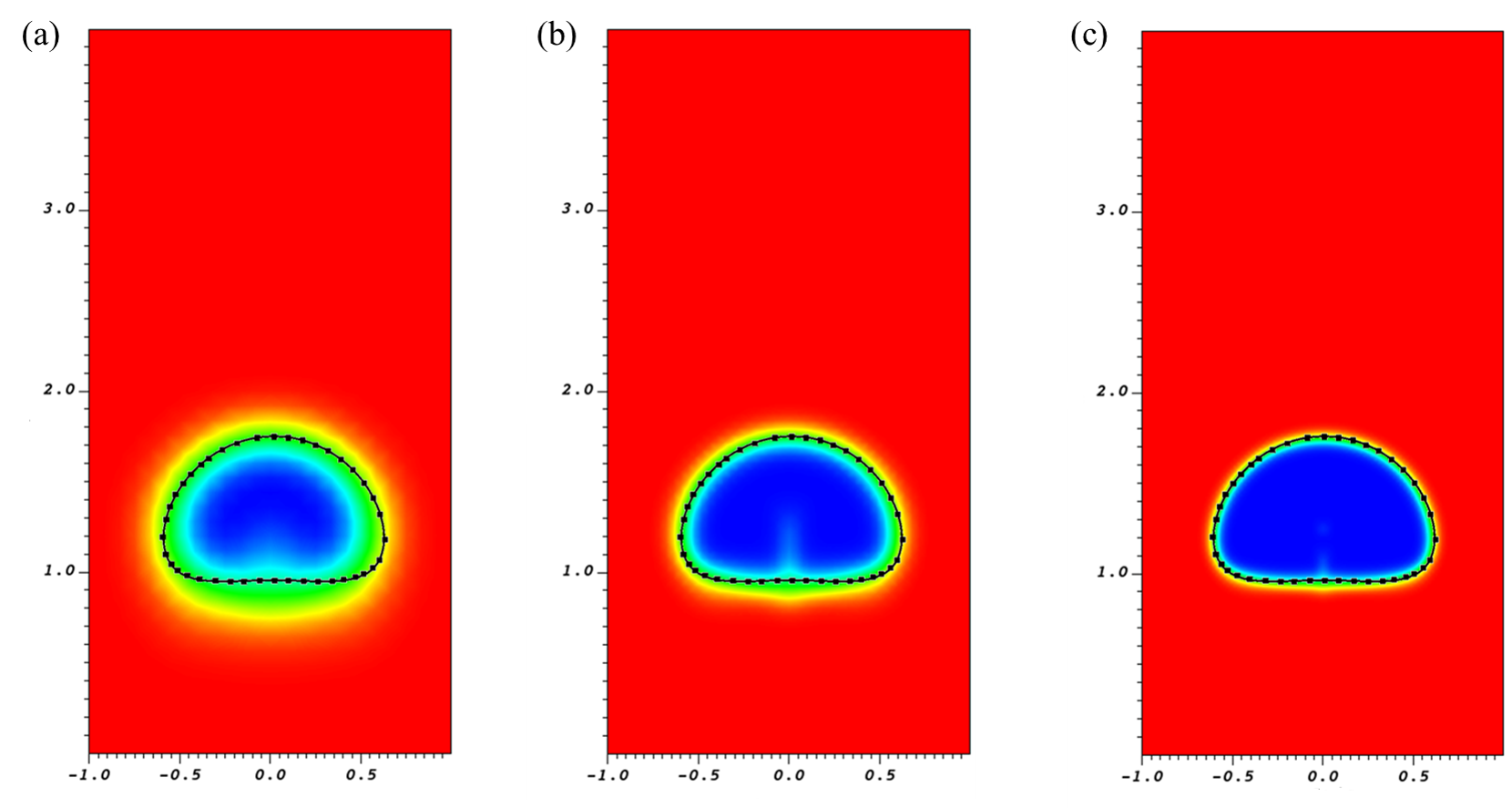}
\caption{Bubble shape comparison at $t = 0.5$: (a) RB1, (b) RB2, (c)RB3. The black filled squares and solid line correspond to the solution of Olsson \citep{olsson2005conservative} and Chiu \citep{chiu2011conservative}.}
\label{Rising_bubble_t_0_5}
\end{figure}

\begin{figure}[htbp]
\centering
\includegraphics[width=0.8\linewidth]{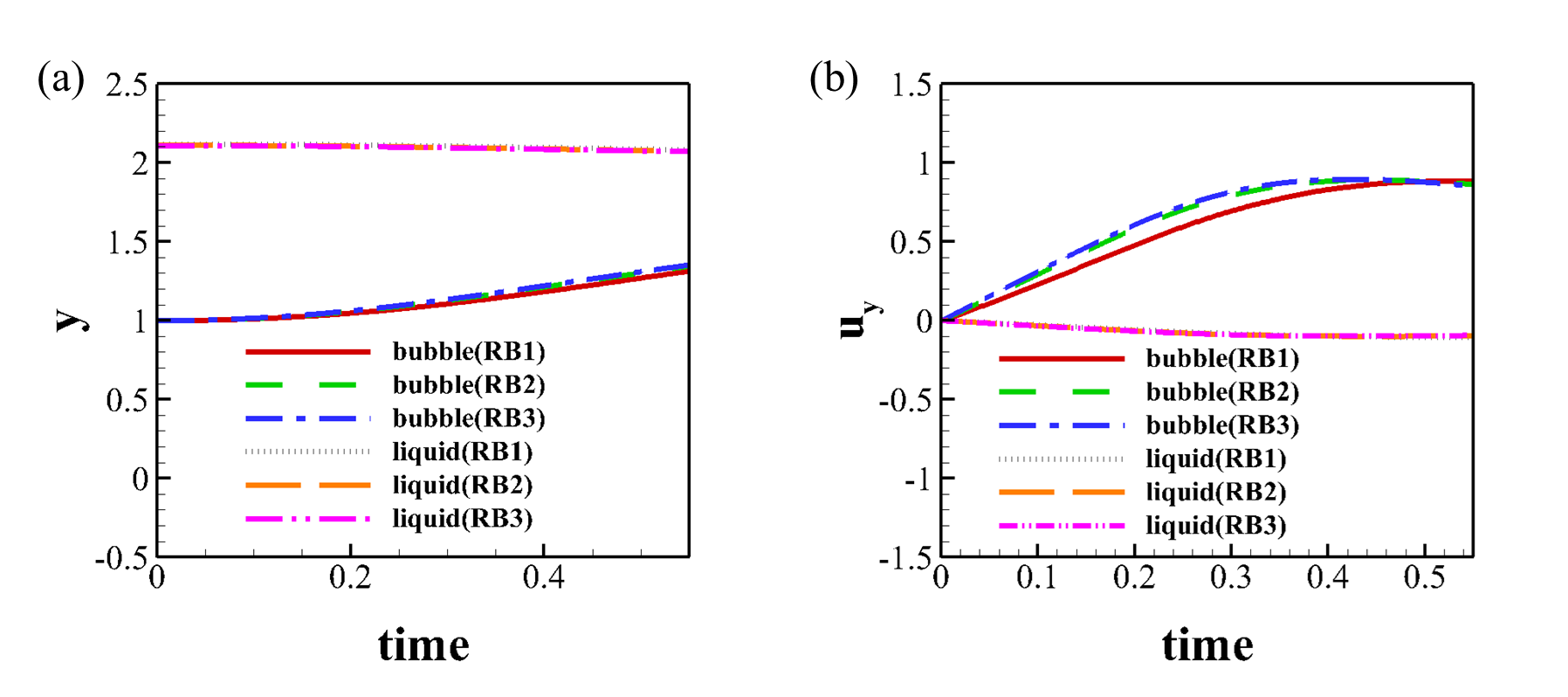}
\caption{The position and velocity of mass center: (a) position of mass center, (b) velocity of mass center. Due to the symmetry (x coordinate of mass center is 0), only the $y-$coordinate is reported here.}
\label{Rising_bubble_mass_center_y_uy}
\end{figure}

Rising bubbles are a frequently occurring phenomenon in the nature and industry.  The simulation of a single rising bubble is a popular  choice for verifying two-phase flow numerical schemes \citep{ding2007diffuse, chiu2011conservative,patel2017novel}. In our numerical experiment, a bubble of dimensionless radius $R=0.5$ in a rectangular domain $[-2R,2R]\times[0,8R]$ is centered at the origin initially with the dimensionless gravity $g_{u}=1.0$. Three regular meshes were used to compute the solution (see Table \ref{Numerical_results_mesh_information_table} for the mesh information). Using the same physical parameters  as that in references \citep{olsson2005conservative,chiu2011conservative}, the dimensionless parameters are: density ratio $\rho_{g}/\rho_{l}$, viscosity ratio $\mu_{g}/\mu_{l}$, Reynolds number $Re$ and Bond number $Bo$ are 0.0013, 0.016, 1111.11 and 3.358, respectively. Here, the subscripts $g$ and $l$ denote the bubble gas and ambient liquid. Also, using the same time scale as that in references  \citep{olsson2005conservative,chiu2011conservative}, the present interface profile at dimensionless time $t=0.5$ (see Figure \ref{Rising_bubble_t_0_5}) compares well with the solution of Olsson \citep{olsson2005conservative} and Chiu \citep{chiu2011conservative}. Due to the symmetry, the bubble and liquid mass center are defined as
\begin{eqnarray}
X_{c}^{bubble}=0, \quad Y_{c}^{bubble}=\frac{\sum\limits_{\sigma^{2}}y(\star\sigma^{2})[1-\phi(\star\sigma^{2})]|\star\sigma^{2}|}{\sum\limits_{\sigma^{2}}[1-\phi(\star\sigma^{2})]|\star\sigma^{2}|}, \quad X_{c}^{liquid}=0, \quad
Y_{c}^{bubble}=\frac{\sum\limits_{\sigma^{2}}y(\star\sigma^{2})\phi(\star\sigma^{2})|\star\sigma^{2}|}{\sum\limits_{\sigma^{2}}\phi(\star\sigma^{2})|\star\sigma^{2}|},
\end{eqnarray}
where $X_c$ and $Y_c$ denote the $x$ and $y$ coordinates of the mass center, respectively. Figure \ref{Rising_bubble_mass_center_y_uy} shows the position and velocity of mass center of bubble and liquid for different meshes, which demonstrates good convergence.


\subsection{Drop impact on a flat pool surface}
\begin{figure}[htbp]
\centering
\includegraphics[width=0.9\linewidth]{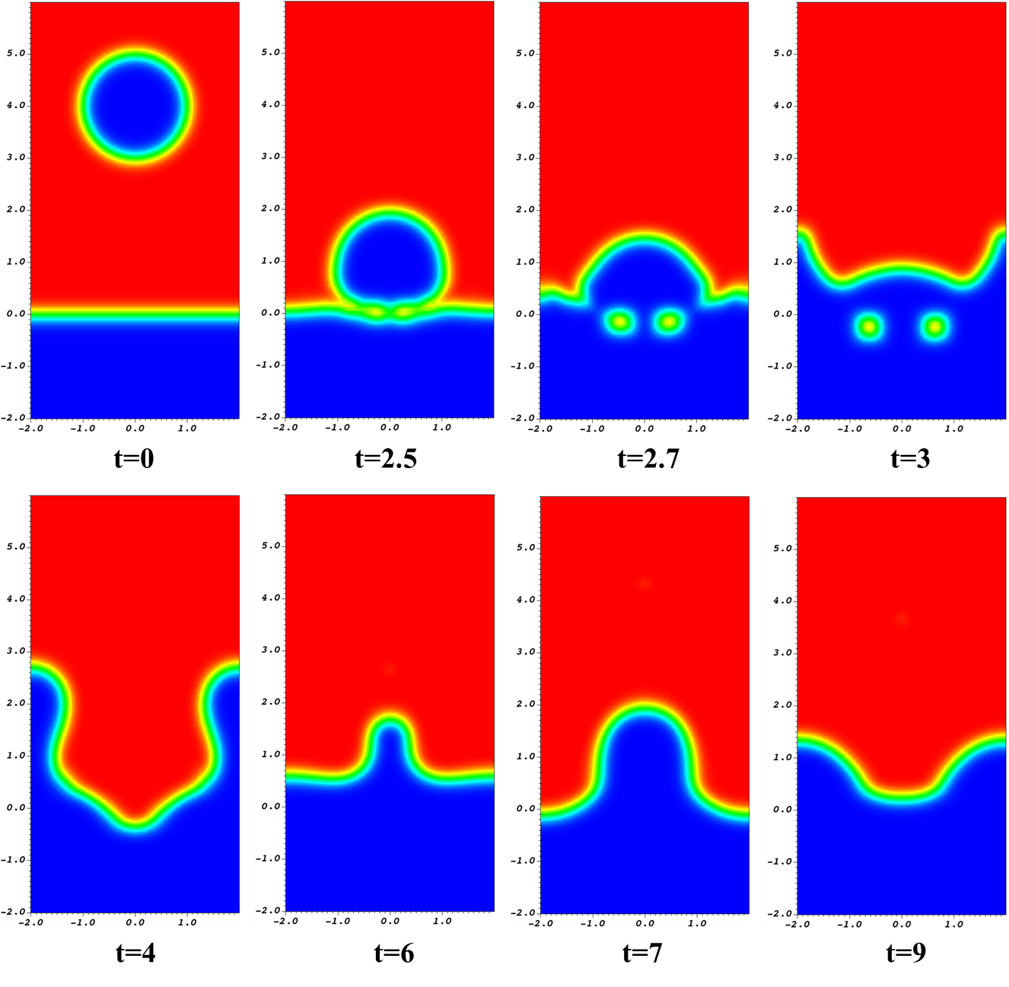}
\caption{The interface evolution for a free falling drop impacting a flat pool surface.}
\label{Drop_pool_evolution}
\end{figure}

The physical phenomenon of a liquid drop falling through the surrounding gas and impacting a flat liquid pool surface involves fascinating interface topological changes. This phenomenon also widely exists in nature, such as rain drops falling on the lake surface. Initially (see $t=0$ frame in Figure \ref{Drop_pool_evolution}), a liquid drop with dimensionless radius $R=1$ in the computational domain $[-2R,2R]\times[-2R,6R]$ is centered at $(0,4R)$ and the surface of the liquid pool is a straight line $y=0$. The mesh information is provided in Table \ref{Numerical_results_mesh_information_table}. We set density ratio $\rho_{r}=\rho_{g}/\rho_{l}=0.001$, viscosity ratio $\mu_{r}=\mu_{g}/\mu_{l}=0.001$, Reynolds number $Re=100$ and Bond number $Bo=1/800$, respectively, and all objects in the computational domain are subjected to the dimensionless gravity $g_{u}=1$. Figure \ref{Drop_pool_evolution} shows the topology change of the interface. The liquid drop is moving downward to the flat pool surface and when the drop touches the pool surface, the coalescence starts. Due to the inertia of the drop, the liquid on both sides is pushed upwards. On the other hand, gravity slows it down and then it falls down gradually. When the falling fluid on each side touch each other, a jet is formed. \textcolor{green}{} Although our results are presently two-dimensional, similar physics of drop impact has been seen in experiments \citep{thoroddsen2000coalescence, thoroddsen2006droplet, kavehpour2015coalescence} for an axisymmetric drop.  An axisymmetrical formulation is currently under development and will be reported in the future with validation against experiments.

\subsubsection{A rising bubble on a curved surface}
\begin{figure}[htbp]
\centering
\includegraphics[width=0.9\linewidth]{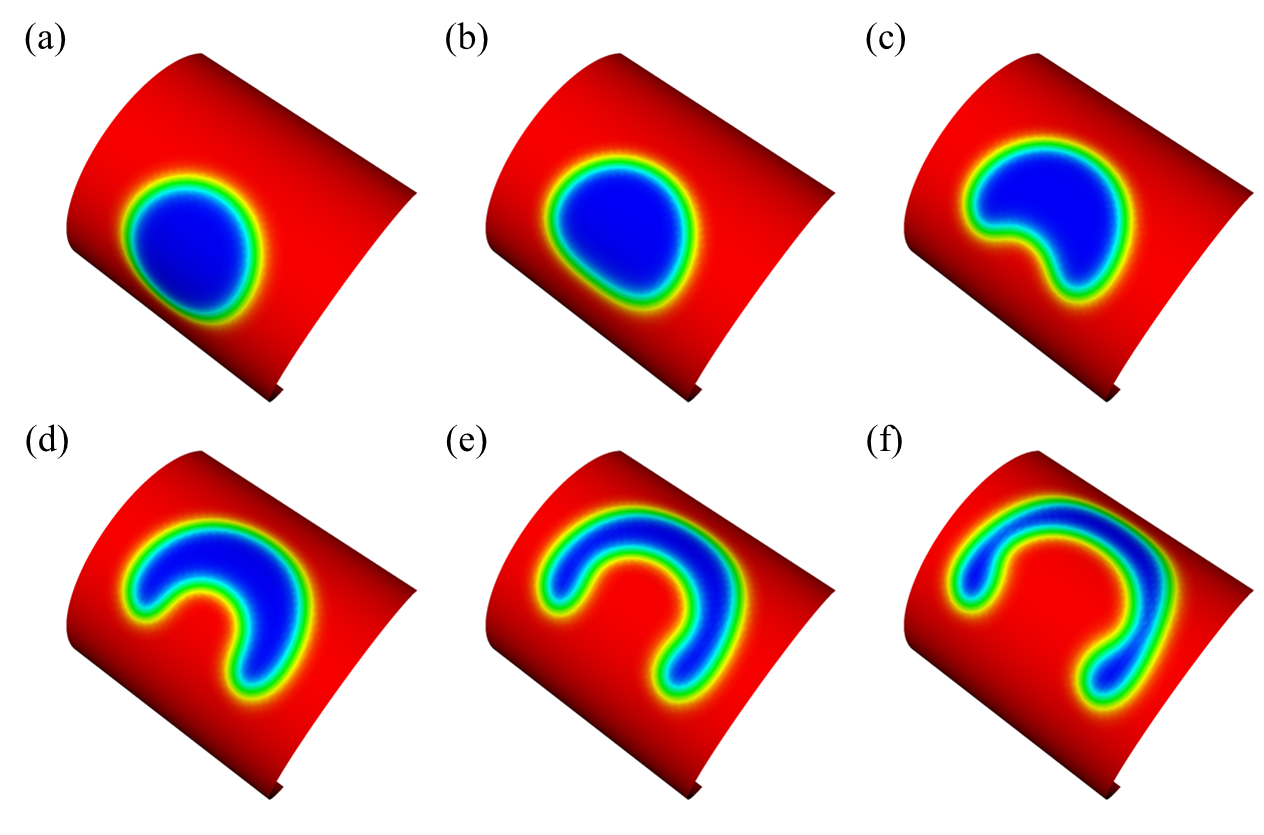}
\caption{The interface evolution of a bubble on a half cylinder: (a) $t=0$, (b) $t=2$, (c) $t=3$, (d) $t=4$, (e) $t=5$, (f) $t=6$.}
\label{Bubble_half_cylinder_evolution}
\end{figure}

\begin{figure}[htbp]
\centering
\includegraphics[width=0.8\linewidth]{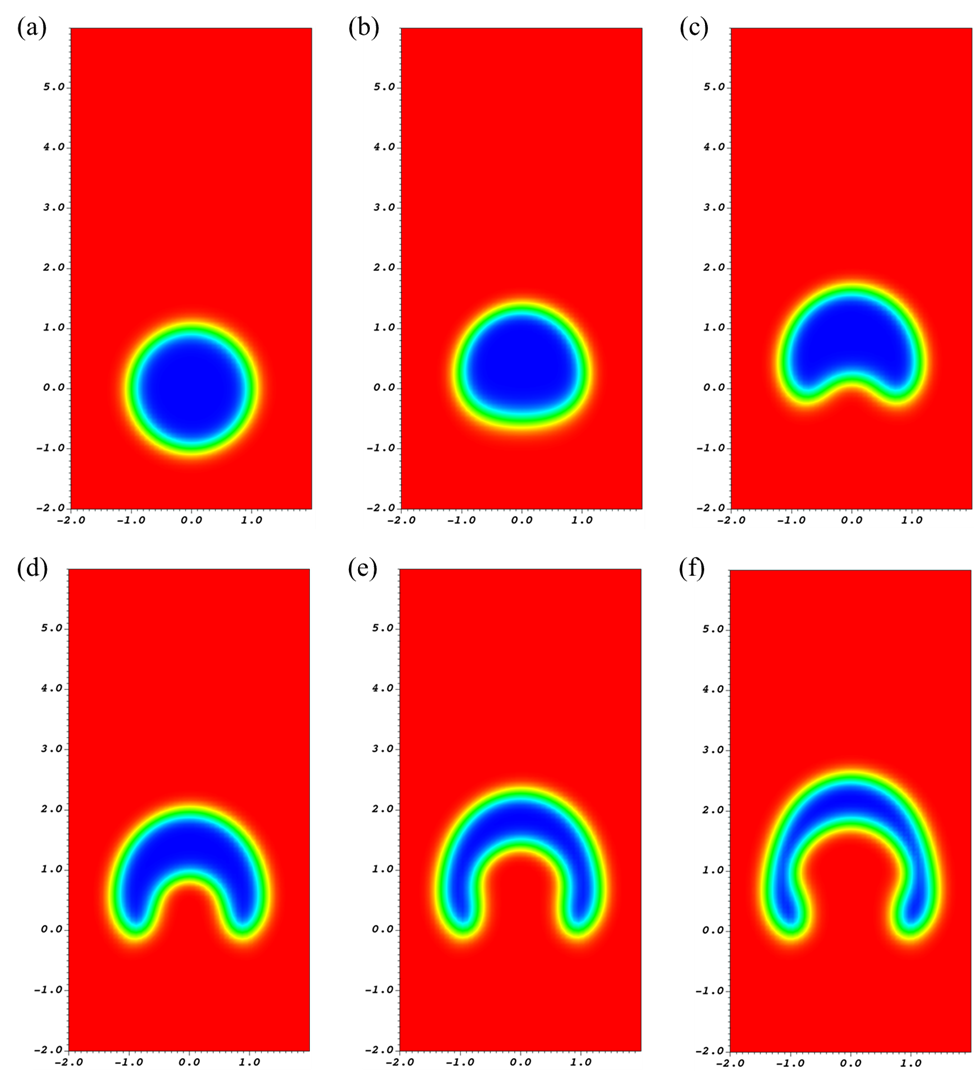}
\caption{The interface evolution of a bubble on a flat plane of size $[-2R,2R]\times[-2R,6R]$: (a) $t=0$, (b) $t=2$, (c) $t=3$, (d) $t=4$, (e) $t=5$, (f) $t=6$.}
\label{Bubble_flat_plane_evolution}
\end{figure}

\begin{figure}[htbp]
\centering
\includegraphics[width=0.9\linewidth]{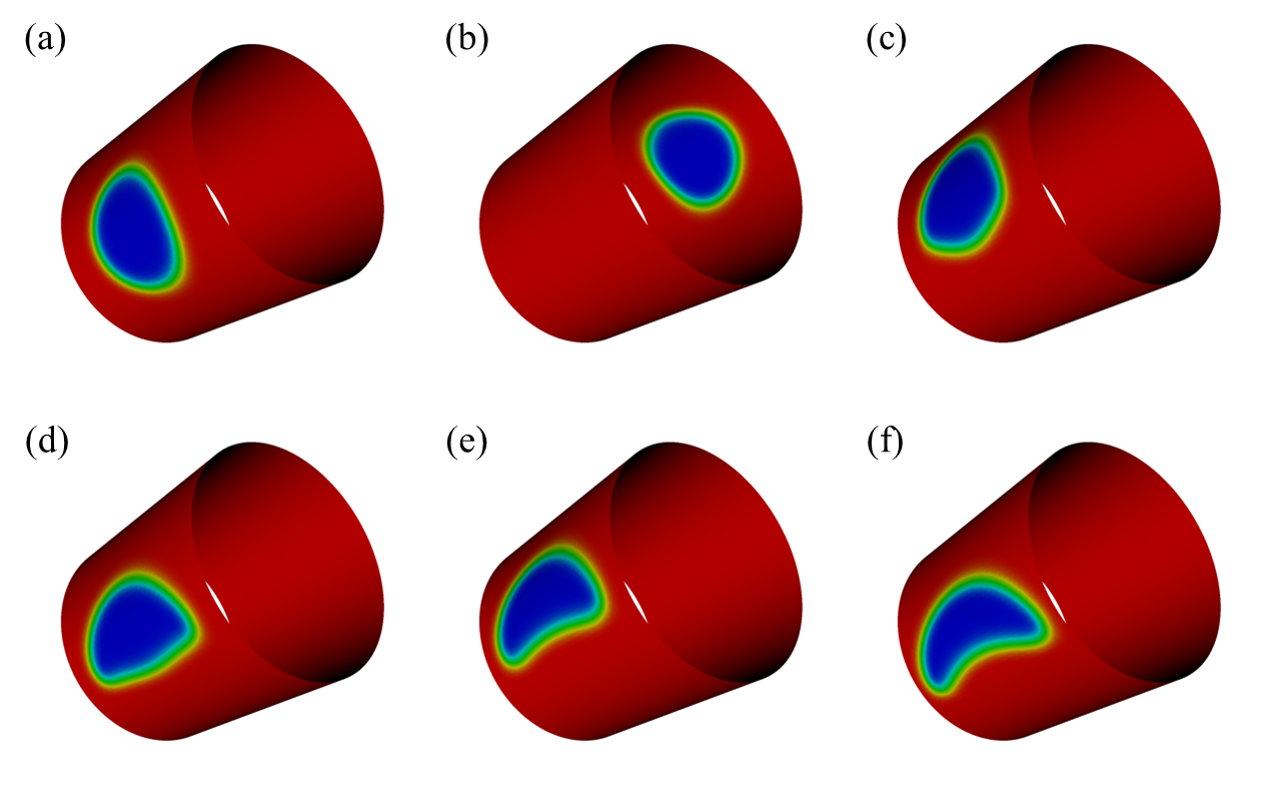}
\caption{The interface evolution of a bubble on a full cylinder: (a) $t=0$, (b) $t=6$, (c) $t=10$, (d) $t=13$, (e) $t=16$, (f) $t=21$.}
\label{Bubble_full_cylinder_evolution}
\end{figure}

DEC operators are independent of the coordinate system employed, therefore it is a convenient method for computing flows on curved surfaces. We consider a rising bubble with initial dimensionless radius $R=1$ embedded on a compact 2D smooth Riemannian manifold and investigate topological change of the interface. Two types of compact 2D Riemannian manifolds, i.e., half cylinder surface (height $h_{h}=4$, radius $r_{h}=2$) and full cylinder surface (height $h_{f}=4$, radius $r_{f}=2$) are used in our simulations, where the subscript $h$ and $f$ denote half and full cylinder surfaces, respectively. Table \ref{Numerical_results_mesh_information_table} shows the mesh information. For both surfaces, the dimensionless gravity $g_{u}=1$ (a body force tangential to the surface in the azimuthal direction),  density ratio $\rho_{r}=0.5$, viscosity ratio $\mu_{r}=1$, Reynolds number $Re=100$, and Bond number $Bo=2500$ are considered. The interface evolution of the bubble on the half cylinder surface (see Figure \ref{Bubble_half_cylinder_evolution}) are identical to that on a flat plane (see Figure \ref{Bubble_flat_plane_evolution}). However, the bubble interface evolution on full cylinder surface (see Figure \ref{Bubble_full_cylinder_evolution}) is quite different from that on a flat plane (see Figure \ref{Bubble_flat_plane_evolution}). Note that the bubble is accelerated continuously because the gravity on full cylinder surface is no longer a conservative force. The bubble flows along the surface with the same interface shape at early time, and with the continuous accelerating effect of the non-conservative gravity force, perceptible bubble interface deformation occurs at late time $t\approx8.7$ (after about three cycles around the cylinder). Exploring the physical dynamics of such deformation is part of our future work.

\section{Summary and future work \label{conclusions}}
We have developed a discrete exterior calculus (DEC) discretization of two phase incompressible Navier-Stokes equations with a conservative phase field method. First, we expressed the governing equations in smooth exterior calculus notation, and then replaced the smooth forms and operators with their discrete counter parts. Here, the dual 1-form at all mesh dual edges, the pressure 0-form at the mesh dual nodes (triangle circumcenters), and the phase-field variable 0-form at the mesh dual nodes are the degrees of freedom. We implement midpoint time integration using a predictor-corrector scheme.   We proved the boundedness of the present method for two time integration schemes, namely the forward Euler and the predictor-corrector time integration scheme in the DEC framework. For a choice of free parameters $\gamma$ and $\epsilon$, the method preserves the boundedness without any ad hoc fixes such as mass redistribution. 

Several verification test cases were presented to compare our simulation results with existing test cases in the literature. We presented several standard advection test cases on planar as well as curved domains, and with regular and irregular meshes, such as reversed single vortex test, Zalesak's disk test and circle translation on a curved surface. These showed that the method preserves mass conservation, and boundedness to machine precision, and the method is convergent with increasing mesh sizes. We then presented results from several test cases two-phase Navier-Stokes simulations. The test cases include the Rayleigh-Taylor instability, dam breaking, rising bubble on flat and curved domains, and drop impact on a pool. Where applicable we have quantitatively verified our results with those in the literature.  The method is able to handle large density and viscosity ratios without any special treatment. Furthermore, the method exhibits mass conservation to the machine precision, exact boundedness, and an error convergence rate between first and second order. 
As part of future work, we will develop an axisymmetric formulation of the multiphase equations. This will facilitate better comparison of the simulation results with the experimental works in the literature. Finally, we expect that an extension to three dimensions to be straightforward, which will make this method a useful tool to investigate multiphase incompressible immiscible flows in 3D complex geometry.

\section*{Acknowledgments}
This research was supported by the KAUST Office of Sponsored Research under Award URF/1/3723-01-01. We thank Nikolas Wojtalewicz (UIUC) for discussions on the boundedness analysis in Section 2.4 for the case of the Euler method. Nikolas Wojtalewicz independently arrived at the same conclusions. We thank Mamdouh Mohamed (Cairo University, Egypt) for discussions.

\bibliographystyle{model1-num-names}
\bibliography{refs}

\end{document}